\documentclass[%
onecolumn,
10pt,
nofootinbib,
 amsmath,amssymb,
 aps,
 a4paper,
 notitlepage,
prd
]{revtex4-1}

\usepackage{graphicx}
\usepackage{dcolumn}
\usepackage{bm}
\usepackage{color}
\usepackage[utf8]{inputenc} 
\begin{document}

\title{Cosmic Acceleration and Growth of Structure in Massive Gravity
}

\author{Michael Kenna-Allison}
\email{michael.kenna-allison@port.ac.uk}
\affiliation{Institute of Cosmology and Gravitation, University of Portsmouth\\ Dennis Sciama
Building, Portsmouth PO1 3FX, United Kingdom}

\author{A. Emir G\"umr\"uk\c{c}\"uo\u{g}lu}
\email{emir.gumrukcuoglu@port.ac.uk}
\affiliation{Institute of Cosmology and Gravitation, University of Portsmouth\\ Dennis Sciama
Building, Portsmouth PO1 3FX, United Kingdom}

\author{Kazuya Koyama}
\email{kazuya.koyama@port.ac.uk}
\affiliation{Institute of Cosmology and Gravitation, University of Portsmouth\\ Dennis Sciama
Building, Portsmouth PO1 3FX, United Kingdom}

\date{\today}

\begin{abstract}
We introduce a cosmological model in the framework of Generalised Massive Gravity. This theory is an extension of non-linear massive gravity with a broken translation symmetry in the St\"uckelberg space. In a recent work, we showed the existence of cosmological solutions stable against linear perturbations. In the present paper, we build up on the previous result and study the evolution of the background solutions and the linear perturbations. At the background level, we find that the mass terms act like a fluid with time dependent equation of state $w < -1$ at late times.
At linear order, we derive the Poisson's equation. We find that the scalar graviton mode invokes anisotropic stress, which brings a modification with respect to LCDM in the effective Newton's constant and the growth rate of matter perturbations. Moreover, we study the propagation of gravitational waves and find that the tensor modes acquire a time dependent mass.
  \end{abstract}
\maketitle


\section{Introduction}

The origin of the late time acceleration of the universe \cite{Perlmutter:1998np,Riess:1998cb} is one of the open questions in cosmology. 
A potential explanation comes from large distance modifications of General Relativity (GR) (see e.g. \cite{Clifton:2011jh, Koyama:2015vza} for reviews).
Massive gravity offers such a solution, however it was not until 2010 when a ghost-free non-linear theory of massive gravity was constructed \cite{PhysRevD.82.044020,PhysRevLett.106.231101} by de Rham, Gabadadze and Tolley (dRGT).

The dRGT massive gravity theory provides the framework for all studies of Lorentz invariant massive gravity. The theory is built out of a \textit{physical metric} $g_{\mu \nu}$, to which matter fields couple minimally, and a \textit{fiducial metric} $f_{\mu \nu}$, which is written in terms of 4 St\"uckelberg fields $\phi^a$ 
introduced to restore covariance. The fiducial metric is defined by
\begin{equation}
    f_{\mu \nu}=\eta_{ab}\partial_{\mu}\phi^a\partial_{\nu}\phi^b\,,
    \label{eq:fiducial-defined}
\end{equation}
where $a,b= 0,1,2,3$ are the field space indices.
The mass term is built out of the elementary symmetric polynomials of $\sqrt{g^{-1}f}$, such that the \textit{Boulware-Deser} ghost, an extra mode that leads to instabilities \cite{PhysRevD.6.3368}, is not introduced. The Minkowski reference metric $\eta_{ab}$ allows the mass term to preserve global Poincar\'e symmetry.

Massive gravity theories allow self-accelerating cosmologies, alleviating the need for a cosmological constant. However, since the introduction of dRGT massive gravity, it has been a challenge to find stable and realistic Friedmann-Lema\^itre-Robertson-Walker (FLRW) solutions in massive gravity \cite{DAmico:2011eto,Gumrukcuoglu:2011ew,Gumrukcuoglu:2011zh,DeFelice:2012mx}, leading to the exploration of extensions of the theory.
One example is bigravity \cite{Hassan:2011zd} where the fiducial metric $f_{\mu \nu}$ is promoted to a dynamical field and comes with its own kinetic term, adding two more degrees of freedom. Other extensions to dRGT with extra scalar fields include \textit{quasidilaton} \cite{DAmico:2012hia}, where the new scalar introduces a conformal factor to the fiducial metric, and mass-varying massive gravity \cite{Huang:2012pe} where the mass parameters are upgraded to functions of the new scalar. Almost all extensions, with varying success in sustaining a stable cosmology, introduce new degrees of freedom in addition to the 5 of dRGT massive gravity. 

Generalised Massive Gravity (GMG) is an extension of dRGT that preserves the number of degrees of freedom, by only breaking the global translation symmetry. This allows the mass parameters to be promoted to functions of invariants of the form $\eta_{ab}\phi^a \phi^b$ \cite{deRham:2014gla}, while preserving Lorentz invariance.\footnote{Starting from constant dRGT terms, the GMG mass terms can be generated through disformal deformations of the reference metric \cite{Gumrukcuoglu:2020utx}.} The theory admits open-FLRW solutions, with all 5 degrees of freedom remaining dynamical throughout the evolution, in contrast with the infinite strong coupling in constant mass dRGT theory \cite{Gumrukcuoglu:2011zh}. The cosmology contains an additional effective fluid with respect to GR, arising from the mass term, which approximates to a cosmological constant. The stability of these backgrounds was shown in some decoupling limit in Ref.\cite{deRham:2014gla} and later in a complete study of linear perturbations \cite{Kenna-Allison:2019tbu}. To our knowledge, this is the only massive gravity theory with 5 degrees of freedom that admits a stable FLRW solution and produces late time acceleration without any other source.

In this paper we build upon the results of Ref.\cite{Kenna-Allison:2019tbu} to follow the evolution of the background and linear scalar perturbations, both analytically and numerically. For concreteness, we focus on a minimal cosmological model from GMG, to obtain numerical predictions for the effective equation of state, the effective Newton's constant and the linear growth of structure. We also study the mass of tensor gravitational waves.

The paper is organised as follows: In Sec.~\ref{GMG} we outline the action of the Generalised Massive Gravity theory. In Sec.~\ref{BGcos} we study the background cosmology, and lay out the steps to obtain solutions for the Hubble rate and the effective equation of state. In Sec.~\ref{pert} we introduce linear scalar perturbations to investigate the effect of the modified background and the scalar graviton mode on the growth of structure. In particular, we identify the scalar graviton from the anisotropic stress, and relate the scalar mode to density contrast. Finally, we obtain the modified Poisson's equation and study the propagation of gravitational waves. In Sec.\ref{summ} we summarise our results and discuss possible future extensions.

\section{Generalised Massive Gravity}\label{GMG}
In this section we review the Generalised Massive Gravity theory.
The gravitational action consists of the Einstein-Hilbert term and the generalised dRGT action \cite{deRham:2014gla}
\begin{equation}\label{model1}
S=\frac{M_p^2}{2}\int d^4 x \sqrt{-g}\left[R+2m^2\sum_{n=0}^4\alpha_n(\phi^a\phi_a)\;\mathcal{U}_n\left[\mathcal{K}\right]\right] +\int d^4 x \sqrt{-g} \mathcal{L}_{matter}\,,
\end{equation}
where $\mathcal{U}_n$ are the dRGT potential terms,
\begin{align}
\mathcal{U}_0(\mathcal{K}) &= 1\,,\nonumber\\
\mathcal{U}_1(\mathcal{K}) &= [\mathcal{K}]\,,\nonumber\\
\mathcal{U}_2(\mathcal{K}) &= \frac{1}{2!}\,([\mathcal{K}]^2-[\mathcal{K}^2])\,,\nonumber\\
\mathcal{U}_3(\mathcal{K}) &= \frac{1}{3!}\,([\mathcal{K}]^3-3[\mathcal{K}][\mathcal{K}^2]+2[\mathcal{K}^3])\,,\nonumber\\
\mathcal{U}_4(\mathcal{K}) &= \frac{1}{4!}\,([\mathcal{K}]^4-6[\mathcal{K}]^2[\mathcal{K}^2]+8[\mathcal{K}][\mathcal{K}^3]+3[\mathcal{K}^2]^2-6[\mathcal{K}^4])\,.
\end{align}
Here, square brackets denote trace operation and the tensor $\mathcal{K}$ is defined by,
\begin{equation}
\mathcal{K}_{\nu}^{\mu}=\delta_{\nu}^{\mu}-\left(\sqrt{g^{-1}f}\right)_{\;\;\nu}^{\mu}\,,
\end{equation} 
where $(\sqrt{g^{-1}f})_{\;\;\rho}^{\mu} (\sqrt{g^{-1}f})_{\;\;\nu}^{\rho}=g^{\mu \rho}f_{\nu \rho}$,
 and the fiducial metric is defined in Eq.\eqref{eq:fiducial-defined}.
In standard dRGT massive gravity, $\phi^a$ are the St\"uckelberg fields, which play the role of restoring general covariance \cite{ArkaniHamed:2002sp}. However, if one abandons 
invariance under translations $\phi^a \to \phi^a +c^a$,
one arrives at the General theory of massive gravity. In this theory, the free parameters $\alpha_n$ of dRGT are promoted to functions of the Lorentz invariant combination $\eta_{ab}\phi^a\phi^b$ \cite{deRham:2014gla}. 

To obtain the equations of motion for the metric, we vary the action (\ref{model1}) with respect to $g^{\mu \nu}$. Performing this operation results in the following gravitational equations of motion,
\begin{equation}\label{coveqns}
    G_{\mu \nu}-\frac{1}{M_p^2}T_{\mu \nu}-m^2\sum_{n=0}^{4}\alpha_n(\phi^a\phi_a)\left(g_{\mu \nu}\;\mathcal{U}_n-2\frac{\delta \mathcal{U}_n}{\delta g^{\mu \nu}}\right)=0,
\end{equation}
where $G_{\mu \nu}$ is the Einstein tensor and $T_{\mu \nu}$ is the energy momentum tensor defined by
\begin{equation}
T_{\mu\nu} \equiv- \frac{2}{\sqrt{-g}}\,\frac{\delta}{\delta g^{\mu\nu}}\left(\sqrt{-g}\,\mathcal{L}_{matter}\right)\,.
\end{equation}
In order to compute the variation of the mass term we define the following tensor,
\begin{align}
X^{\alpha}_{\;\;\beta} \equiv \left(\sqrt{g^{-1}f}\right)^\alpha_{\;\;\beta}, \quad 
X^\alpha_{\;\;\beta}X^\beta_{\;\;\gamma} = g^{\alpha\beta}f_{\beta\gamma}\,.
\end{align}
Using this definition, we can vary the trace of various powers of this tensor:
\begin{equation}
\delta [X^n] = \frac{n}{2}\,(X^n)^\alpha_{\;\;\mu}\;g_{\alpha\nu}\delta g^{\mu\nu}\,,
\end{equation}
which is valid for any power $n\ge 1$. The variation of the mass terms can therefore be written in the following form \cite{Babichev:2013pfa,Kenna-Allison:2018izo}
\begin{align}
\label{variation}
\frac{\delta \mathcal{U}_1}{\delta g^{\mu\nu}} &= -\frac{1}{2}\, X^\alpha_{\;\;\mu} g_{\alpha\nu}\,,\nonumber\\
\frac{\delta \mathcal{U}_2}{\delta g^{\mu\nu}} &= \left[\left(-\frac{3}{2}+\frac{1}{2}[X]\right)X - \frac{1}{2} X^2 \right]^\alpha_{\;\;\mu}
g_{\alpha\nu}\,,\nonumber\\
\frac{\delta \mathcal{U}_3}{\delta g^{\mu\nu}} &= \left[\left(-\frac{3}{2}+[X]-\frac{1}{4}[X]^2+\frac{1}{4}[X]^2 \right)X + \left(-1+\frac{1}{2}[X]\right) X^2 -\frac{1}{2}X^3\right]^\alpha_{\;\;\mu}
g_{\alpha\nu}\,,\nonumber\\
\frac{\delta \mathcal{U}_4}{\delta g^{\mu\nu}} &= \left[\left(-\frac{1}{2}+\frac{1}{2}[X]-\frac{1}{4}[X]^2+\frac{1}{12}[X]^3+\frac{1}{4}[X^2] -\frac{1}{4}[X][X^2]+\frac{1}{6}[X^3] \right)X \right.\nonumber\\
& \left. \qquad+ \left(-\frac{1}{2} +\frac{1}{2}[X]-\frac{1}{4}[X]^2+\frac{1}{4}[X^2]\right) X^2 +\left(-\frac{1}{2}+\frac{1}{2}[X]\right)X^3-\frac{1}{2}X^4\right]^\alpha_{\;\;\mu}
g_{\alpha\nu}\,.
\end{align}

For the matter sector, we assume a source which satisfies the covariant conservation law
\begin{equation}
 \nabla^\mu T_{\mu\nu} = 0\,.
 \label{eq:mattereom}
\end{equation}
With a conserved source, the equations for the St\"uckelberg fields are automatically implied by the metric equations, through the contracted Bianchi identity
\cite{Hassan:2011vm}
\begin{equation}
 \nabla^\mu \left(\frac{2}{\sqrt{-g}} \frac{\delta S}{\delta g^{\mu\nu}}\right) = \frac{1}{\sqrt{-g}}\,\frac{\delta S}{\delta \phi^a}\,\partial_\nu \phi^a\,.
 \label{eq:stuckeq}
\end{equation}

\section{Background Cosmology}\label{BGcos}

In this section we study the background cosmology of the GMG theory.
\subsection{Set up}
In order to preserve homogeneity and isotropy, we require that \textit{(i)} $\eta_{ab}\phi^a\phi^b$ is uniform; \textit{(ii)} $g$ and $f$ have the same FLRW symmetries such that no coordinate dependence arises from $g^{-1}f$. Both of these requirements can be satisfied only if we consider an open FLRW background \cite{deRham:2014gla,Kenna-Allison:2019tbu}
\begin{equation}\label{gmetric}
    g_{\mu \nu}dx^{\mu}dx^{\nu}=-dt^2+a(t)^2 \Omega_{ij}dx^i dx^j\,,
\end{equation}
$\Omega_{ij}$ is the metric of the constant time hypersurfaces with constant negative curvature
\begin{equation}
\Omega_{ij}dx^i dx^j=dx^2+dy^2+dz^2-\frac{\kappa (xdx+ydy+zdz)^2}{1+\kappa(x^2+y^2+z^2)}\,,
\end{equation}
and $\kappa = |K| = -K$ is the absolute value of the negative constant curvature of the spatial slice. 
For this background, the unique St\"uckelberg field configuration that satisfies the homogeneity and isotropy conditions is \cite{Gumrukcuoglu:2011ew}
\begin{align}\label{fields}
\phi^0&=f(t)\sqrt{1+\kappa(x^2+y^2+z^2)}\,,\nonumber\\
\phi^1&=f(t)\sqrt{\kappa}x \,,\nonumber\\ 
\phi^2&=f(t)\sqrt{\kappa}y \,,\nonumber\\
\phi^3&=f(t)\sqrt{\kappa}z \,.
\end{align}
With this definition, the fiducial metric corresponds to Minkowski space-time in an open chart
\begin{equation}\label{f} 
f_{\mu \nu}dx^{\mu}dx^{\nu}=-\dot{f}(t)^2dt^2+\kappa f(t)^2 \Omega_{ij}dx^i dx^j\,,
\end{equation}
where an overdot denotes time derivative.%

For the matter sector we consider a perfect fluid described by the energy momentum tensor,
\begin{equation}\label{energymom}
    T_{\mu \nu}=\rho \,u_{\mu}u_{\nu}+P(g_{\mu \nu}+u_{\mu}u_{\nu}),
\end{equation}
where $u_{\mu}$ is the 4-velocity of the fluid which satisfies the normalisation condition $u_{\mu}u^{\mu}=-1$. For the rest of the work we will restrict our analysis to a non-relativistic fluid corresponding to $P=0$.

\subsection{Background Dynamics}\label{bg}
To obtain the background equations, we substitute the metric ansatze (\ref{gmetric}) and (\ref{f}) into the equations of motion (\ref{coveqns}) and \eqref{eq:mattereom}. Doing so results in the following background equations,
\begin{align}\label{BGeqns}
3\,\left(H^2-\frac{\kappa}{a^2}\right)&=m^2L+\frac{\rho}{M_p^2}\,,\nonumber\\
2\left(\dot{H}+\frac{\kappa}{a^2}\right)&=m^2J(r-1)\xi-\frac{\rho}{M_p^2} \,,\nonumber\\
\dot{\rho}&=-3\,H\,\rho\,,
\end{align}
where for convenience, we defined
\begin{equation}
H \equiv \frac{\dot{a}}{a} \,,\qquad 
\xi \equiv\frac{\sqrt{\kappa}f}{a}\,,\qquad 
r\equiv \frac{a\,\dot{f}}{\sqrt{\kappa}f}\,.
\label{eq:defHXr}
\end{equation}
We also define three combinations of the mass 
functions for ease of notation,
\begin{align}
\label{defLJ}
L & \equiv -\alpha_0+(3\,\xi-4)\alpha_1-3\,(\xi-1)(\xi-2)\alpha_2+(\xi-1)^2(\xi-4)\alpha_3+(\xi-1)^3\alpha_4
\,,\nonumber\\
J&\equiv\alpha_1+(3-2\xi)\alpha_2+(\xi-1)(\xi-3)\alpha_3+(\xi-1)^2\alpha_4\,,\nonumber\\
\Gamma &\equiv  \xi J+\frac{(r-1)\xi^2}{2}J'(\xi),
\end{align}
where $ \alpha_n = \alpha_n(\phi^a \phi_a)$. For the field configuration (\ref{fields}), $\phi^a\phi_a \to -f(t)^2$ which ensures homogeneity and isotropy. From the Friedmann equation, we identify  $m^2M_p^2L$ as the effective energy density arising from the mass term, while from the acceleration equation we recognise $m^2M_p^2J(1-r)\,\xi$ as the sum of the effective energy density and pressure. Using the contracted Bianchi identities \eqref{eq:stuckeq} we can obtain the background St\"uckelberg equation,
\begin{equation}\label{stuck}
3\,H\,J\,(r-1)\xi-\dot{L}=0\,,
\end{equation}
or equivalently,
\begin{equation}\label{stuck-alt}
3\,\left(H-\frac{\sqrt{\kappa}}{a}\right)\,J=-\frac{2\,a\,\xi}{\sqrt{\kappa}}
\left[
-\alpha'_0+(3\,\xi-4)\alpha'_1-3\,(\xi-1)(\xi-2)\alpha'_2+(\xi-1)^2(\xi-4)\alpha_3'+(\xi-1)^3\alpha_4'\right]\,.
\end{equation}

From (\ref{stuck-alt}) it is clear to see the advantages of Generalised Massive Gravity over dRGT massive gravity. In the dRGT limit, the right hand side vanishes, which either forces $J=0$, leading to an infinite strong coupling problem \cite{Gumrukcuoglu:2011zh}; or $H=\frac{\sqrt{\kappa}}{a}$ which does not allow for expansion \cite{DAmico:2011eto,Gumrukcuoglu:2011zh}. As shown in \cite{Kenna-Allison:2019tbu}, the generalised case evades the strong coupling problem of dRGT 
and admits a stable cosmology. To determine the background evolution, we study in more detail the Friedmann and St\"uckelberg equations, then derive the Hubble rate and the equation of state of dark energy. In order to gain an analytic understanding of the model, we work within the minimal model of GMG as introduced in \cite{Kenna-Allison:2019tbu}, where the $\alpha$-functions take the following form:
\begin{align}
\alpha_0(\phi^a\phi_a)&=\alpha_1(\phi^a\phi_a)=0\,,\nonumber\\
\alpha_2(\phi^a\phi_a)&=1+m^2\alpha_2' \phi_a\phi^a\,,\nonumber\\
\alpha_3(\phi^a\phi_a)&=\alpha_3 \,,\nonumber\\
\alpha_4(\phi^a\phi_a)&=\alpha_4 \,.
\label{alphas}
\end{align}
For $\alpha_2'\ll 1$, the contribution $m^2M_p^2 L$ to the Friedmann equation becomes approximately a cosmological constant, and its time dependence is controlled by the dimensionless parameter $\alpha_2'$. 

Assuming that $|\phi^a\phi_a|=f(t)^2$ increases with time, this minimal model asymptotes to dRGT at early times, although the strong coupling problem is still tamed with respect to dRGT \cite{Kenna-Allison:2019tbu}. With this choice, the Friedmann and St\"uckelberg equations take the following form:
\begin{align}\label{mini}
&3\left(H^2-\frac{\kappa}{a^2}\right)-m^2\left[-3(\xi-2)(\xi-1)(1-m^2f^2\alpha_2')+\alpha_3(\xi-1)^2(\xi-4)+\alpha_4(\xi-1)^3\right]=\frac{\rho}{M_p^2}\,,\nonumber\\
&3\left(H-\frac{\sqrt{\kappa}}{a}\right)\left[(3-2\xi)(1-m^2f^2\alpha_2')+\alpha_3(\xi-3)(\xi-1)+\alpha_4(\xi-1)^2\right]=
6\,m^2f\,(\xi-2)(\xi-1) \alpha_2'\,.
\end{align}
In order to investigate the mass term's effective equation of state and the Hubble rate, we first introduce dimensionless variables and re-write every dimensionful quantity in 
units of $H_0$ and $M_p$
\begin{align}
m &\to H_0 \mu\,,\nonumber\\
H &\to H_0 h\,,\nonumber\\
\kappa &\to a_0^2H_0^2\Omega_{\kappa 0} \,,\nonumber\\
\alpha_2' &\to \frac{q}{\mu^2}\,,\nonumber\\
\rho &\to \frac{3a_0^3H_0^2M_p^2\Omega_{m 0}}{a^3}\,.
\end{align}
With this choice the parameter $q$ now controls how far away from dRGT we are, where $q \to 0$ is the dRGT limit. We also use Eq.\eqref{eq:defHXr} to replace $f(t)$ with $f(t) \to \frac{a \xi}{\sqrt{\kappa}}$. With these replacements, (\ref{mini}) becomes:
\begin{align}
\label{mini-dimless1}
&3\left(h^2-\frac{a_0^2\Omega_{\kappa 0}}{a^2}\right)-\mu^2\left[
\frac{3\,(\xi-2)(\xi-1)(q a^2\xi^2-a_0^2\Omega_{\kappa 0})}{a_0^2\Omega_{\kappa 0}}
+\alpha_3(\xi-1)^2(\xi-4)+\alpha_4(\xi-1)^3\right]=\frac{3a_0^3\Omega_{m 0} }{a^3} \,,\\
\label{mini-dimless2}
&3\left(h-\frac{\sqrt{\Omega_{\kappa 0}}a_0 }{a}\right)\left[(3-2\xi)
\left(1-\frac{q \,a^2 \xi^2}{a_0^2\Omega_{\kappa 0} }\right)
+\alpha_3(\xi-3)(\xi-1)+\alpha_4(\xi-1)^2 \right]=\frac{ 6q a(\xi-2)(\xi-1)\xi}{a_0 \sqrt{\Omega_{\kappa 0}}}\,.
\end{align}
The full evolution for $h$ and $\xi$ in terms of $a$ can be determined by solving the above equations. 

For numerical solutions, we will fix the following parameters \footnote{Note we will allow $\Omega_{\kappa 0}$ to vary later in this section.}:

\begin{equation}
\Omega_{\kappa 0} = 3\times 10^{-3},\qquad\Omega_{m 0} = 0.3,\qquad \alpha_3 = 0,\qquad \alpha_4 = 0.8\,,
\label{eq:parameters}
\end{equation}
where the specific choice of $\alpha_n$ parameters correspond to a simple choice within the allowed parameter space for stable cosmologies in Ref.\cite{Kenna-Allison:2019tbu}.

\subsection{Varying q}
In this section we study the background cosmology of GMG, whilst varying the parameter $q$ which controls the deviation away from dRGT. 

From here on we rescale q in the following way
\begin{equation}
Q\equiv 10^{4}q\,,
\end{equation}
for the purpose of clarity later in the plots.

First, we outline the method taken to isolate the physical solution for $\xi(a)$ and $h(a)$. Initially, we keep $\mu$ arbitrary since it is sensitive to the value of $Q$ and will later be fixed by requiring that the effective energy density from the mass term is consistent with the choice of cosmological parameters. We first solve the St\"uckelberg equation (\ref{mini-dimless2}) for $h(\xi,a,Q,\mu)$. We then replace this solution in the Friedmann equation \eqref{mini-dimless1}, which results in a 10th order polynomial equation for $\xi(a,Q,\mu)$. 
To choose the physical solution with positive real values, we compare the values of the roots of this equation to the value of $\xi_{dRGT}$ at early times, where the contribution from $\alpha_2'$, or $Q$, is negligible. We set aside the roots that are closest. The solution for $\xi$ in dRGT is \cite{Gumrukcuoglu:2011ew},
\begin{equation}
    \xi_{dRGT}^{\pm}=\frac{1+2\alpha_3 \pm\sqrt{1+\alpha_3+\alpha_3^2-\alpha_4}+\alpha_4}{\alpha_3+\alpha_4},
\end{equation}
which arises from solving $J(\xi)=0$, i.e. the quadratic equation (\ref{stuck-alt}) when the right hand side is zero. As shown in Ref.~\cite{Kenna-Allison:2019tbu}, only the 
$+$ root
allows for a real tensor mass so we work with this solution. Using the parameter values \eqref{eq:parameters}, the solution is $\xi_{dRGT}^+ = 2.80902$. We can then compare the solutions for $\xi_{GMG}$ to this value. Fig.~\ref{xiplot} shows the solution for $\xi_{GMG}$ in comparison to $\xi_{dRGT}^+$:
\begin{figure}[ht]
    \begin{center} 
    \includegraphics[width=0.7\columnwidth]{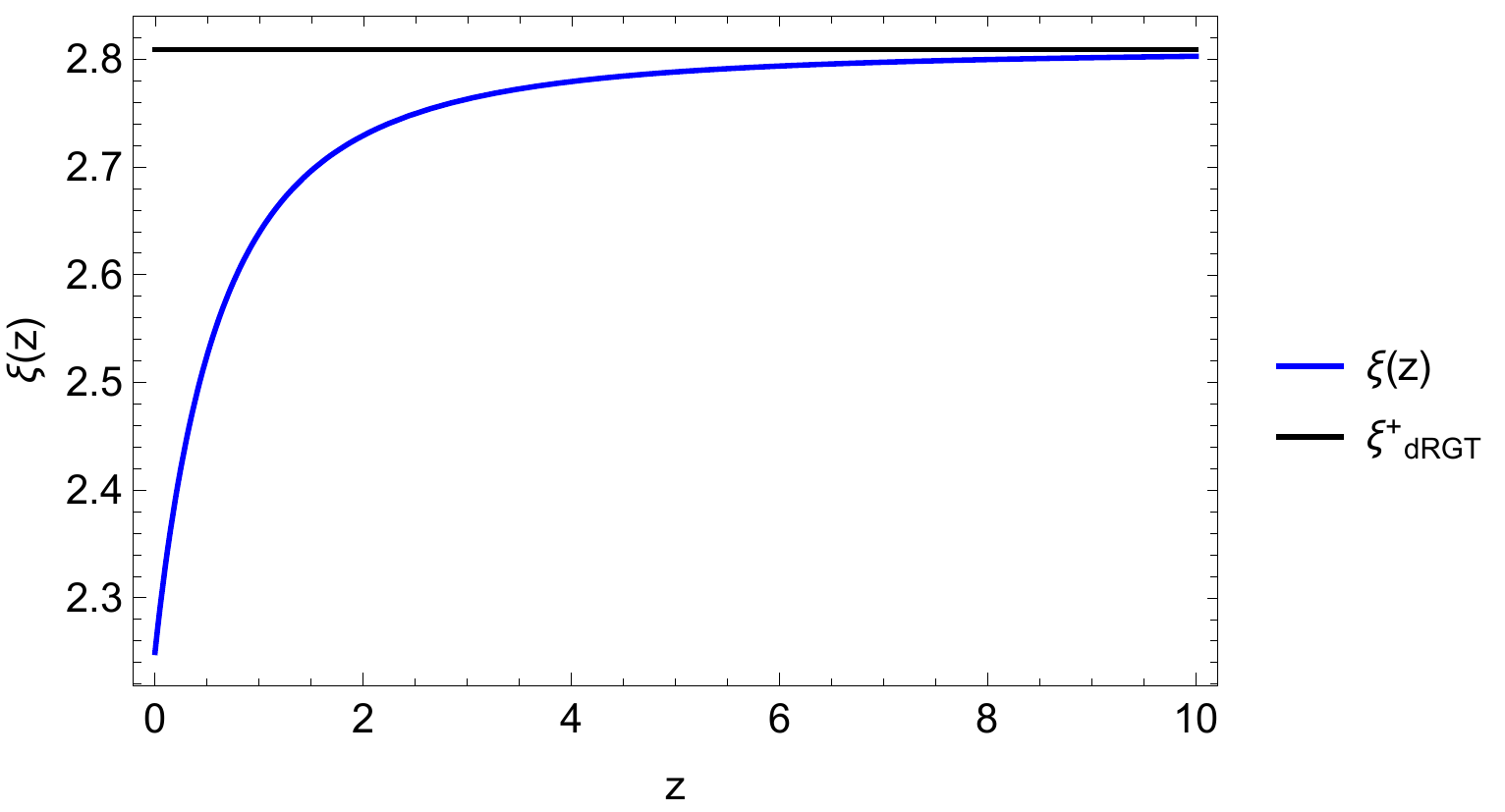}

    \end{center}
    \caption{The black solid line shows the solution for $\xi_{dRGT}$. The blue curve shows the only physical solution for $\xi_{GMG}$ which asymptotes to the solution for $\xi_{dRGT}$ at early times. Parameter values taken are $Q=1$ and those outlined in \ref{bg}.}
 
    \label{xiplot}
\end{figure}
8 of the other solutions do not converge to $\xi_{dRGT}^+$ at early times, whilst the other $\xi$ solution which tends to $\xi_{dRGT}^+$ at early times does not satisfy both (\ref{mini-dimless1}) and (\ref{mini-dimless2}), leaving us with one physical solution.

We have reduced the system to one $\xi(a,Q,\mu)$ solution and it's corresponding $h(a,Q,\mu)$ solution. We start by first determining the value of $\mu$ that would be compatible with the cosmological parameters today. We rewrite the Friedmann equation Eq.\eqref{mini-dimless1} as,
\begin{equation}
 h^2 - \frac{a_0^2\,\Omega_{\kappa 0}}{a^2}  = \Omega_{DE} +\frac{a_0^3\,\Omega_{m 0}}{a^3}\,,
 \label{fried1}
\end{equation}
where we defined the density function for the effective dark energy as
\begin{equation}
\Omega_{DE} \equiv \frac{\rho_{DE}}{3\,H_0^2\,M_p^2} = \frac{\mu^2\,L}{3}
=\frac{\mu^2}{30}\,(\xi-1)\left[8\,(\xi-1)^2+(\xi-2)\left(\frac{Q\,\xi^2\,a^2}{a_0^2}-30\right)\right]\,.
\label{lambda}
\end{equation}
For a given value of $Q$, we evaluate this equation today $a=a_0$ using the  root $\xi(a_0,Q,\mu)$. Since today we have $\Omega_{\kappa 0}+\Omega_{m0}+\Omega_{DE 0}=1$, we fix the value of $\mu$ using this relation.

We now investigate the effect of varying the parameter $Q$ on the expansion rate and the equation of state of the effective fluid for the mass term. The effective equation of state can be obtained by identifying the contribution $(P^{DE}+\rho^{DE})$ from the acceleration equation, i.e. the second line of Eq.~(\ref{BGeqns}) 
\begin{equation}
     P^{DE}+\rho^{DE} = - M_p^2 m^2 J(r-1)\xi\,,
 \end{equation}
then using the effective density defined in Eq.\eqref{lambda}. As a result, we find
\begin{equation}
w_{DE}= \frac{P_{DE}}{\rho_{DE}} = -1-\frac{\mu^2 J (r-1)\xi}{3\,\Omega_{DE}}= -1 -\frac{J (r-1)\xi}{L}\,.
\label{w}
\end{equation}
The functions $J$ and $L$, defined in \eqref{defLJ}, are functions of $\xi$ and $a$ only, so using the solution $\xi(a)$, we can determine their evolution with $a$. For the quantity $r$, we use Eq.\eqref{eq:defHXr} to write
\begin{equation}
r = \frac{a}{\sqrt{\kappa}}\,\left(H + \frac{\dot\xi}{\xi}\right)\,,
\label{eq:defr}
\end{equation}
which can be calculated by using the solutions $h$, $\xi$ and its derivative.

\begin{figure}[ht]
    \begin{center} 
    ~\!\!\!\!\!\!\!\!\!\!\!\!\!\!\!\includegraphics[width=0.5\columnwidth]{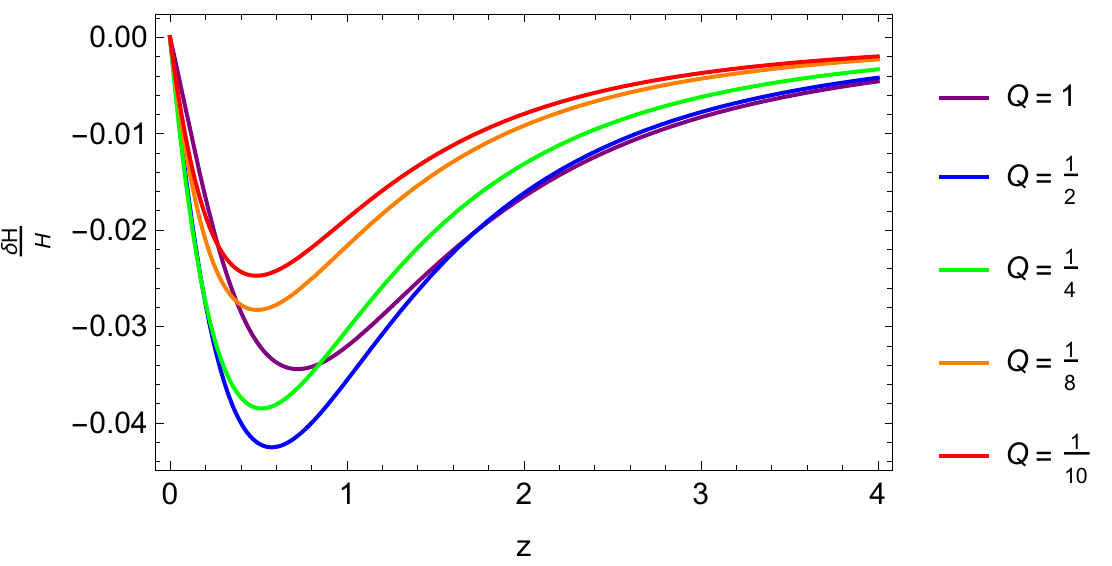}~~~~~~~~
\includegraphics[width=0.5\columnwidth]{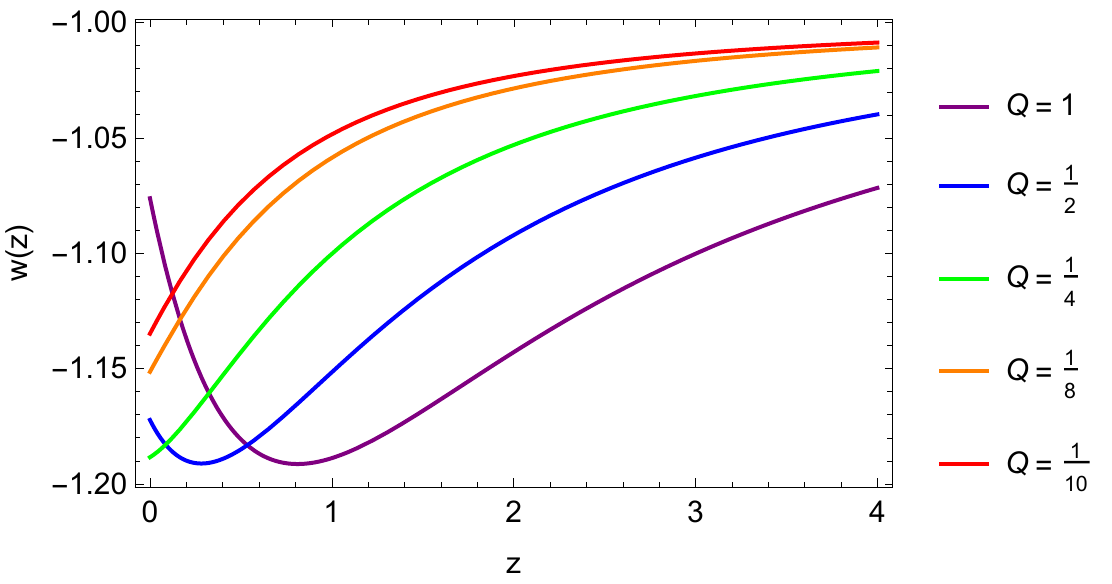}
    \end{center}
    \caption{Left panel shows fractional deviation in the Hubble rate in GMG compared to LCDM with varying values of $Q$, where $\frac{\delta h}{h}=\frac{h-h_{LCDM}}{h_{LCDM}}$.  The right panel shows the equation of state $w(z)$.}
    \label{varyingq}
\end{figure}

We can now discuss the effect of $Q$ on the evolution. In Fig.\ref{varyingq} we show the effect of varying $Q$ on the Hubble rate and the equation of state.

Smaller values of $Q$ lead to smaller deviations of the Hubble rate from the LCDM value, which are typically at 1\% level. This is due to $Q$ controlling how close we are to the dRGT background where the effective density from the mass term is constant. We note that the normalisation of $\mu$ parameter ensures that the value of $H$ recovers the LCDM value today. The evolution of the equation of state parameter shows a late time departure from a cosmological constant, with maximum deviation at a 20\% level with $w_{DE}<-1$, followed by a bounce back towards $w_{DE}=-1$. Notably, the value of $Q$ does not affect the size of this departure but only the time it occurs.
This behaviour can be understood as follows. The departure from dRGT is introduced in $\alpha_2$ via Eq.\eqref{alphas}
\begin{equation}
    \alpha_2=1-\frac{Q}{10^4}f^2,
\end{equation} 
Since $f$ is an increasing function of time, with $f^2 \propto (1+z)^{-2}$ a smaller value of $Q$ will simply decrease the redshift where the deviation from dRGT starts and delay the bounce in $w_{DE}$ to later times. Conversely, a large value of $Q$ will cause the bounce at an earlier time, thus allowing $w_{DE}$ to increase back up and cross $w_{DE}=-1$. From \eqref{w}, this crossing occurs when either $J\sim0$ or $r\sim 1$. When $J$ crosses zero, the solutions collapse to the self-accelerating solutions of dRGT, and scalar and vector modes becomes infinitely strongly coupled \cite{Gumrukcuoglu:2011zh}. On the other hand, there is no a priori reason that prevents $r$ from crossing $1$. In the next section, we will determine the conditions for which the background solution exists and how, if at all, it breaks down. 

To study the regime of applicability of the solution, we will impose the perturbative stability conditions derived in Ref.\cite{Kenna-Allison:2019tbu}. We summarise them in Appendix \ref{A}. We then evolve the background equations until one of the above conditions are broken, which then results in the theory no longer being applicable as a dark energy model.

\subsection{Case Study: $Q=1$}
In this section we give a concrete example of the cosmological evolution for a value of $Q=1$ with $\Omega_{\kappa 0} = 3\times10^{-3}$. We classify the evolution into several points.

\begin{enumerate}
    \item The evolution starts at $z=10$, at early times the solution tracks the dRGT evolution.
    \item The equation of state undergoes a decrease away from $w=-1$ reaching a minimum value of $w \sim -1.19$ at $z \sim 0.8$, and there is a small deviation away from the LCDM value shown in the Hubble rate at about 2\% level.
\item $w$ then starts to bounce back to $w=-1$ and the difference in the Hubble rate decreases as we evolve towards $z=0$ and into the future, past $z=0$.
\item The critical point in the evolution now occurs. At $z_{c} \sim -0.178 $, $w$ crosses $-1$ which is caused by $r=1$. At this point $\Gamma$ also crosses $0$ which generates a tachyonic instability in the tensor sector \eqref{eq:stability-tensor}. This is not problematic though as the instability takes the age of the universe to develop as the mass is typically of order Hubble. 
However, as the sound speeds of the vector modes are also proportional to $\Gamma$ \eqref{eq:stability-vector} and $J$ is still positive, 
the vector modes become unstable and the background solution is no longer valid.
\end{enumerate}
We see that $Q=1$ marginally misses the point of instability. Requiring that the solution does not break down before $z=0$, we will only consider the values $Q \leq 1$.

\subsection{Varying $\Omega_{\kappa 0}$}
\begin{figure}[ht]
    \begin{center} 
    ~\!\!\!\!\!\!\!\!\!\!\!\!\!\!\!\includegraphics[width=0.5\columnwidth]{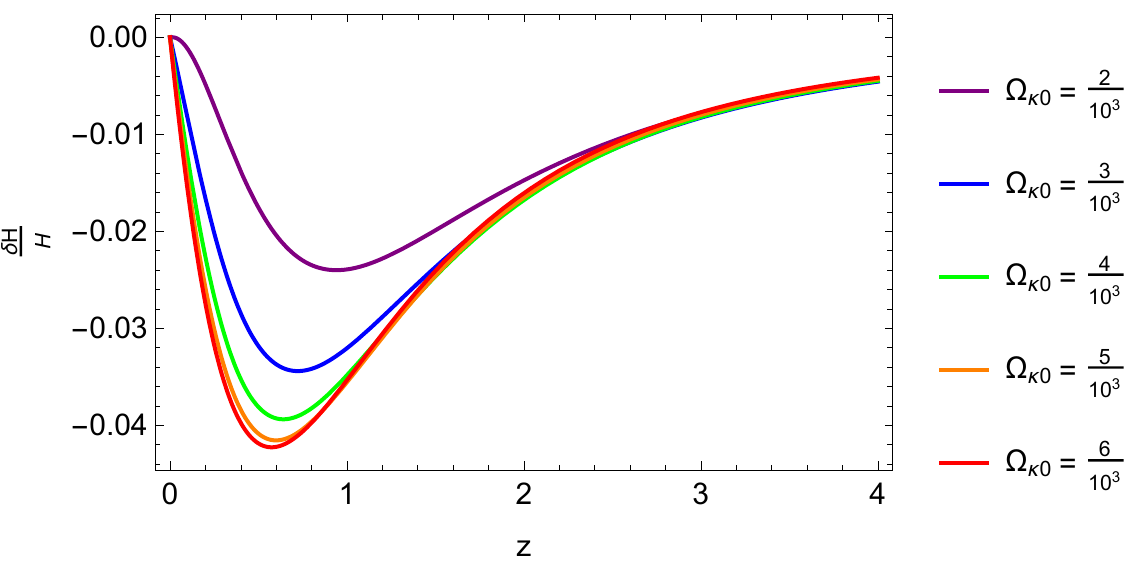}~~~~~~~~
\includegraphics[width=0.5\columnwidth]{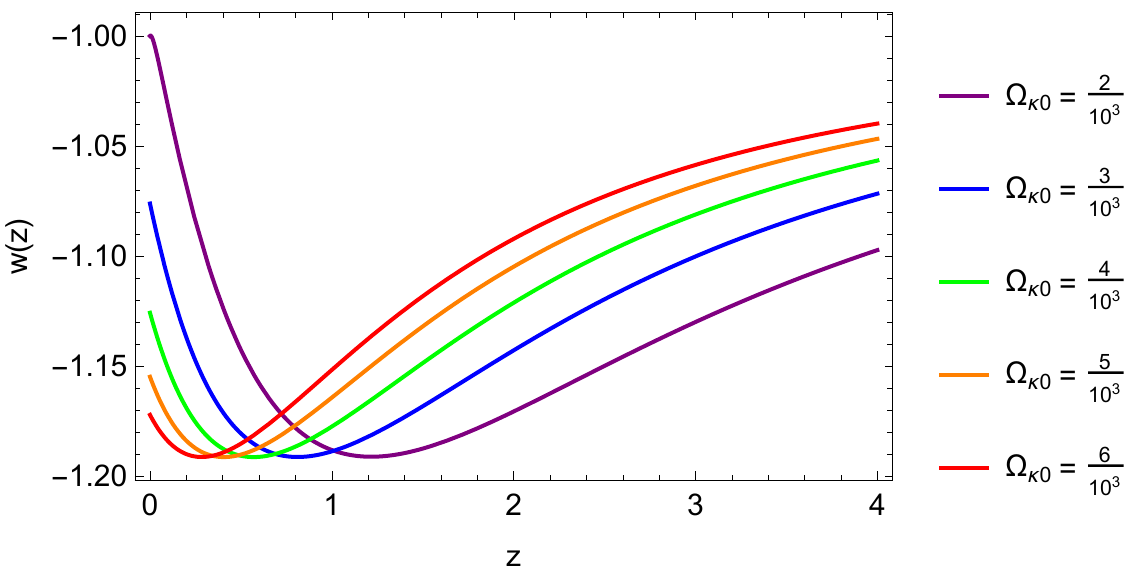}
    \end{center}
    \caption{For different values of $\Omega_{\kappa0}$, left panel shows fractional deviation in the Hubble rate in GMG compared to LCDM, $\frac{\delta h}{h}=\frac{h-h_{LCDM}}{h_{LCDM}}$, and the right panel shows the equation of state. 
}
    \label{varyingkappa}
\end{figure}

The evolution of the universe is also sensitive to the value of $\Omega_{\kappa 0}$. As can be seen from (\ref{mini-dimless1}), shifting the value of $\Omega_{\kappa 0}$, under the condition that 
$\Omega_{\kappa 0}+\Omega_{m 0}+\Omega_{DE 0}=1$,
effectively shifts the value of 
 $\Omega_{DE 0}$,
whilst keeping $\Omega_{m 0}$ fixed. 
Fixing $Q=1$, we plot the Hubble rate and the equation of state in Fig.~\ref{varyingkappa}.

Notably, a higher value of $\Omega_{\kappa 0}$ pushes the time the instability occurs further into the future. The fractional deviation in the Hubble rate increases for higher values of $\Omega_{\kappa 0}$. For a value of $\Omega_{\kappa 0}=6\times10^{-3}$, the deviation is around 4\% at $z \approx 0.8$. There are two partnering effects one can see here. For a fixed value of $\Omega_{\kappa 0}$, decreasing $Q$ pushes the time the instability is generated into the future. Furthermore, for a fixed $Q$ increasing $\Omega_{\kappa 0}$ also pushes the time the instability is generated into the future. 

To wrap up, we have outlined a case study of the cosmological evolution in the minimal model of GMG. We find that the expansion history of the Universe 
can be matched for values of $Q \leq 1$. For a value of $Q=1$, we can describe the evolution up to a redshift of $z \approx -0.2$ before a gradient instability is generated in the vector sector. Decreasing the value of $Q$ or increasing $\Omega_{\kappa 0}$, pushes the time the instability is generated further into the future. 
The equation of state undergoes a bounce from $w<-1$ to $w=-1$ and has a dynamical dark energy--like effect which could be constrained using observations.

\section{Linear Perturbations}\label{pert}
In this section we outline the study of linear perturbations and the derivation of the Poisson's equation to determine the effect of GMG on the growth of structure.
\subsection{Set up}
Only the linear scalar perturbations are relevant for the derivation of the Poisson's equation, 
so we decompose the g-metric takes as:
\begin{equation}\label{decomp1}
g_{\mu \nu}dx^{\mu}dx^{\nu}=-(1+2\phi)dt^2+a\,\partial_i B \,dt\,dx^i+a^2\left(\Omega_{ij}+h_{ij}\right)dx^i dx^j,
\end{equation}
with $h_{ij}$ decomposed as
\begin{equation}\label{decomp}
h_{ij}=2\,\psi\, \Omega_{ij}+\left(D_i \,D_j-\frac{1}{3}\Omega_{ij}D_lD^l\right)S\,.
\end{equation}
In this set up, $D_l$ is the covariant derivative compatible with the 3-metric $\Omega_{ij}$. Spatial indices are raised and lowered with $\Omega^{ij}$. Perturbations to the matter sector are introduced via $\rho(t,x^i) = \rho(t)+\delta \rho(t,x^i)$ and $u^{\mu}=(1-\phi,\partial^i v)$, where $v$ is the longitudinal component of the velocity perturbation, this leads to the following form for $T^{\mu}_{\;\; \nu}$:
\begin{align}\label{tmunu}
T^0_{\;\; 0}&=-(\delta \rho +\rho)\,,\nonumber\\
T^0_{\;\; i}&=a \rho(\partial_i B+a\partial_i v)\,,\nonumber\\
T^i_{\;\; 0}&=
-\rho\partial^i v\,,
\nonumber\\
T^i_{\;\; j}&=0\,.\nonumber\\
\end{align}
We exhaust the gauge freedom by fixing the unitary gauge, $\delta \phi^a = 0$, thus the fiducial metric $f_{\mu\nu}$ remains unperturbed. Using the metric decomposition \eqref{decomp1}, along with Eqs.\eqref{f} and \eqref{tmunu}, we calculate the perturbed Einstein's equations.
Schematically, we end up with 4 coupled Einstein's equations denoted by $\mathcal{E}^{00},\mathcal{E}^{0i},\mathcal{E}^{tr},\mathcal{E}^{tl}$,
where ``tr'' and ``tl'' denote the trace and traceless parts of the $(ij)$ equation, respectively.
The conservation of the energy-Momentum tensor, $\nabla_{\mu}T^{\mu}_{\;\;\nu}$ $=0$ yields the Euler $\mathcal{E}^{eu}$ and continuity $\mathcal{E}^{co}$ equations for the matter fluid. 
The full expressions of the equations of motion for linear perturbations are summarised in appendix~\ref{app:einsteineqs}.

\subsection{Linear Perturbation Analysis}
The unitary gauge that we are using leaves the metric and matter perturbations intact. In order to compare with GR, it is useful to define gauge invariant variables that can be constructed out of the available fields. We start with the coordinate transformation
\begin{equation}\label{GT}
    x^{\mu} \to x^{\mu}+\delta x^{\mu},
\end{equation}
where $\delta x^{\mu}= (\delta x^0 , \Omega^{ij} \partial_j \delta x)$ is of order of perturbations. Under \eqref{GT}
the metric and matter perturbations transform as,
\begin{align}
    \phi& \to \phi+\delta \dot x^0\,,\nonumber\\
    B & \to B + a\,{\delta \dot x}- \frac{1}{a}\,\delta x^0\,,\nonumber\\
    \psi & \to \psi +\frac{1}{3} D_iD^i \delta x+H\,\delta x^0\,,\nonumber\\
    S & \to S + 2\,\delta x\,,\nonumber\\
    \delta \rho &\to \delta \rho + \dot\rho \,\delta x^0\,,\nonumber\\
    v & \to v - \delta \dot x\,.
\end{align}
With these transformations, we can construct the gauge-invariant variables that are analogues of the Newtonian gauge in GR:
\begin{align}\label{GI1}
    \Phi&=\phi-\partial_t \zeta\,,\nonumber\\
    \Psi&=\psi-H\zeta-\frac{1}{6}D^iD_i S \,,\nonumber\\
    \tilde{\delta \rho}&=\delta \rho-\dot{\rho}\,\zeta \,,\nonumber\\
    \tilde{v}&=v+\frac{1}{2}\dot{S},
\end{align}
where
\begin{equation}
    \zeta\equiv -a\,B+\frac{1}{2}a^2 \dot{S}.
\end{equation}
We also define a convenient combination for density contrast
\begin{equation}
\Delta \equiv \frac{\tilde{\delta\rho}}{\rho} - 3\,a^2\,H\,\tilde{v}\,.
\end{equation}

We then substitute (\ref{GI1}) into the equations of motion (\ref{coveqns}) and expand all perturbation variables in scalar harmonics,
\begin{align}
    \Psi &=\int dk\; k^2 \Psi_{|\vec{k}|} \,Y (\vec{k},\vec{x}) \nonumber\\
    S&= \int dk \; k^2  S_{|\vec{k}|} \,Y(\vec{k},\vec{x}) \nonumber\\
    &\vdots
\end{align}
where the scalar harmonics satisfy 
$D^iD_i Y \to -k^2 Y$, 
to yield the first order perturbation equations outlined in appendix \ref{app:einsteineqs}. Looking at the form of the traceless equation $\mathcal{E}^{tl}$, we see that $S$ acts as a source of anisotropic stress. In GR, the equation for the anisoptropic stress vanishes, ie: $\Phi+ \Psi =0$. Whilst in GMG, we have
\begin{equation}
\Phi+ \Psi =\frac{1}{2}m^2 a^2 \Gamma S.
\end{equation}
This reveals that the dynamical scalar degree of freedom $S$ could mediate a fifth force which would alter the structure of the Poisson's equation. Our goal is to investigate if $S$ affects structure formation, and what scales the modifications to gravity appear at due to the presence of this extra mode. To do this, we first derive the equation of motion for $S$. We start by solving the matter equations for $\dot{\tilde{v}}$ and $\dot{\Delta}$,
\begin{equation}\label{matsol}
    \dot{\tilde{v}}=-\frac{2a^2H\tilde{v}+\Phi}{a^2} \,,\qquad \dot{\Delta}=3H\Phi+\tilde{v}\Bigg[k^2+3\kappa+\frac{3}{2}m^2a^2(1-r)J\xi+\frac{3a^2 \rho}{2M_p^2}\Bigg]-3\dot{\Psi},
\end{equation}
and substitute it into the equations of motion to reduce the system to the 4 Einstein's equations. We next solve the constraint equations $\mathcal{E}^{00}$ and $\mathcal{E}^{0i}$ for the non-dynamical degrees of freedom $\Phi$ and $B$,
\begin{align}\label{phib}
    \Phi&=\frac{1}{12M_p^2a^2 H^2}\bigg[-4(k^2+3\kappa)M_p^2r\Psi-3a^4H(2\tilde{v}\rho+m^2M_p^2J\xi \dot{S})+a^2\Big(k^2m^2Jr\xi S+2r\rho \Delta -6m^2M_p^2 J \xi r\Psi+12M_p^2H\dot{\Psi}\Big)\Bigg]\,\nonumber\\
    B&=\frac{r+1}{6m^2M_p^2a^3HJr\xi}\bigg[-a^2(m^2M_p^2J\xi(k^2 S-6\Psi)+2\rho \Delta)+4(k^2+3\kappa)M_p^2 \Psi+3m^2M_p^2a^4HJ\xi\dot{S}\bigg].
\end{align}
Upon substituting (\ref{phib}) and the time derivatives $(\dot{\Phi}, \dot{B})$ into $\mathcal{E}^{tr}$ and $\mathcal{E}^{tl}$, we end up with two equations $\mathcal{E}^{tr}$ and $\mathcal{E}^{tl}$ in terms of $(\Psi, S, \tilde{v})$ and their time derivatives. We then solve $\mathcal{E}^{tr}$ for $\Psi(S,\dot{S},\ddot{S},\tilde{v},\Delta)$ and replace into $\mathcal{E}^{tl}$ as well as the time derivative $\dot{\Psi}$. Upon doing this, we obtain a second order differential equation for $S$ sourced by matter perturbations $(\Delta, \tilde{v})$.

\begin{equation}\label{master}
   \mathcal{A}\ddot{S}+H_0 \mathcal{B}\dot{S}+H_0^2\mathcal{C}S+\mathcal{D} \Delta +H_0\mathcal{F}\tilde{v}=0.
\end{equation}
The dimensionless coefficients $(\mathcal{A},\mathcal{B},\mathcal{C},\mathcal{D},\mathcal{F})$ are functions of $(k,z)$ and various background quantities relating to the GMG theory. It is pertinent to note that no assumptions have been made at this stage so this equation is the most general equation one can write down for the scalar $S$.

In order to gain an analytic understanding of the system, we adopt the minimal model \eqref{alphas} once more, in which we expand each of the background GMG quantities around their dRGT values as below:
\begin{align}
J &= \frac{Q}{10000}J_1 + \mathcal{O}\left(\frac{Q}{10000}\right)^2\,,\nonumber\\
\Gamma&=\Gamma_0+ \frac{Q}{10000} \Gamma_1 +\mathcal{O}\left(\frac{Q}{10000}\right)^2 \,, \nonumber\\
r & = \frac{a\,H}{\sqrt{\kappa}} +\frac{Q}{10000} r_1 +\mathcal{O}\left(\frac{Q}{10000}\right)^2 \,, \nonumber\\
\end{align}
where the leading order terms $J_1$ and $\Gamma_0$ are not necessary for this discussion and can be found in Ref.\cite{Kenna-Allison:2019tbu}. 

\begin{figure}[ht]
 \begin{center} 
    ~\!\!\!\!\!\!\!\!\!\!\!\!\!\!\!\includegraphics[width=0.5\columnwidth]{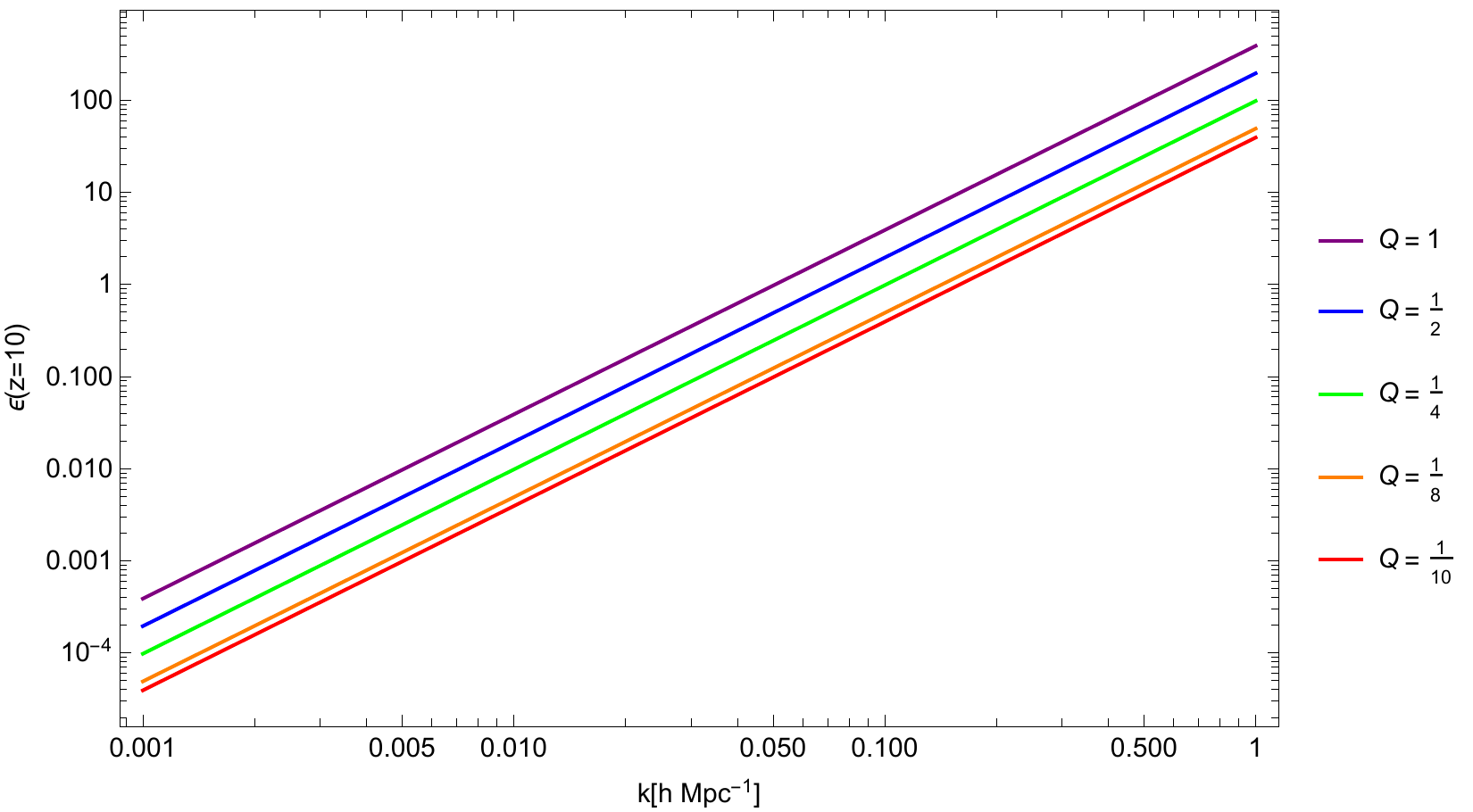}~~~~~~~~
\includegraphics[width=0.5\columnwidth]{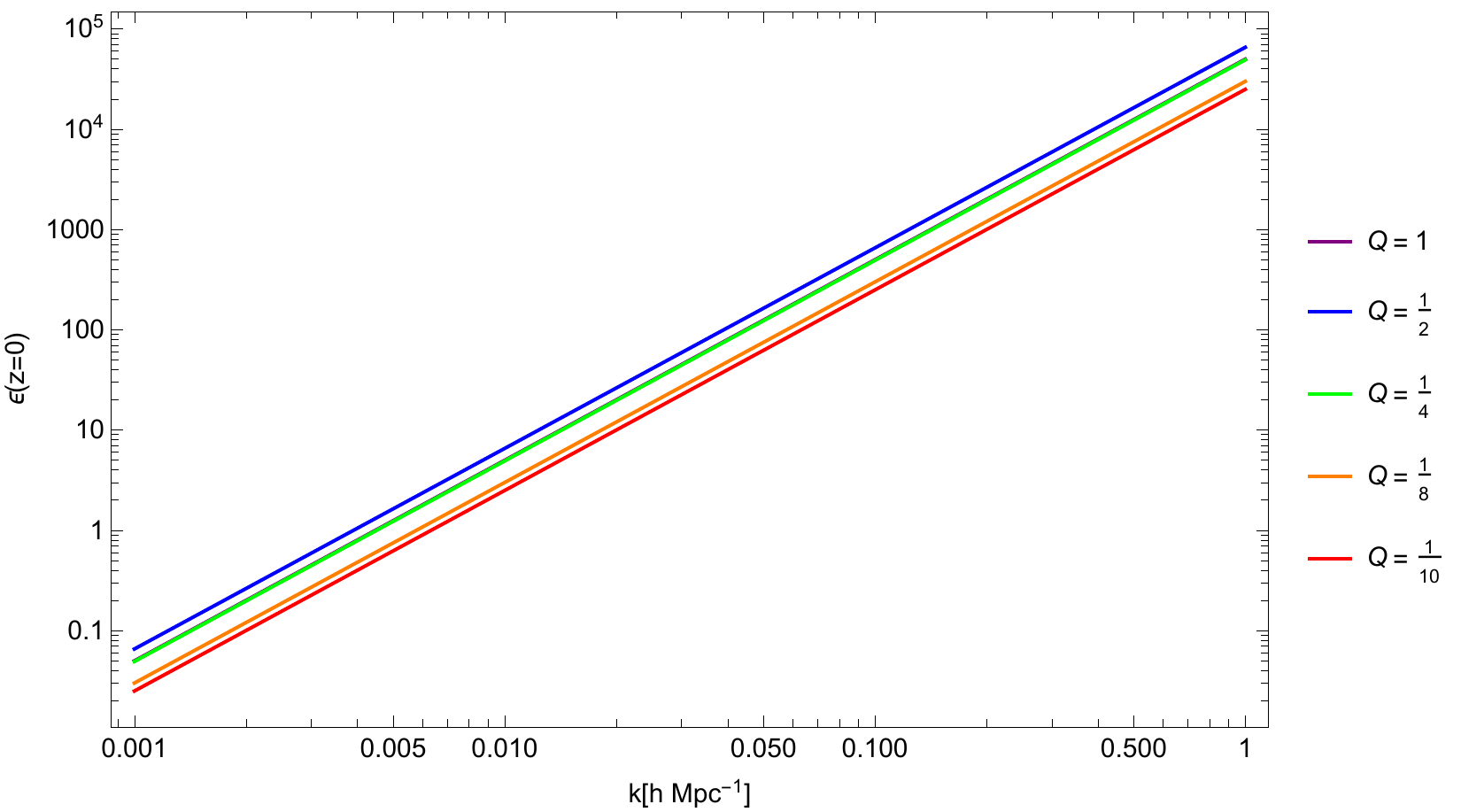}
    \end{center}
    \caption{Right panel shows $\mathcal{E}$ evaluated at $z=0$, whilst the left panel is $z=10$. }
    \label{varyingeps}
\end{figure}

A peculiar feature of the scalar terms in the UV is the appearance of a new energy scale $H/J$ in addition to the scale of expansion $H$. For small departures from dRGT we have $J <1$. Depending on the relation to these scales, the subhorizon modes $k\ll a\,H$ can have two distinct behaviours. We define the following dimensionless parameter to distinguish between the cases \cite{Kenna-Allison:2019tbu}
\begin{equation}
 \mathcal{E} \equiv \frac{k^2J}{a^2H^2}\,.
\end{equation}
There are three qualitatively distinct limits:
\begin{enumerate}
 \item $1 \ll \mathcal{E} \ll \frac{k^2}{a^2H^2}$ : subhorizon modes with momenta larger than the new scale $H/J$. For these modes, the variation of the mass parameters due to GMG affect the UV behaviour. These cases are captured by first performing the subhorizon (UV) expansion, then the dRGT expansion.
 \item $\mathcal{E} \ll 1 \ll \frac{k^2}{a^2H^2}$ : subhorizon modes with momenta smaller than the new scale. The leading order terms in the UV expansion for these modes are identical to dRGT, although the deviation due to GMG does not lead to infinite strong coupling. These cases correspond to first carrying out the dRGT expansion, then the subhorizon expansion.
 \item $\mathcal{E} \ll \frac{k^2}{a^2H^2}\ll 1 $ : superhorizon modes. There is no ambiguity for this case, and it corresponds to taking the superhorizon limit.
\end{enumerate}
For varying values of $Q$, and for the parameters given in Eq.~\eqref{eq:parameters}, we plot $\mathcal{E}(k)$ for $z=10$ and $z=0$. The result is shown in Fig. (\ref{varyingeps}).
$\mathcal{E}$ being larger for a specified value of $k$ at $z=0$ is expected as $\mathcal{E} \propto J$:
at later times the departure from the early dRGT (or LCDM) behaviour, controlled by $J$, increases. 

This feature is also seen in Fig.\ref{varyingq}, deviations away from LCDM occur at late times, whilst at early times LCDM is recovered. At $z=10$, larger values of $Q$ favour a larger $\mathcal{E}$ however at $z=0$ this is not the case: the $Q=1$ line coincides with the $Q=1/4$, whilst $Q=1/2$ gives the largest value of $\mathcal{E}$. This feature corresponds to the bounce of the effective equation of state observed in Fig.\ref{varyingq}.

\begin{figure}[htp]
\centering
\includegraphics[width=.3\textwidth]{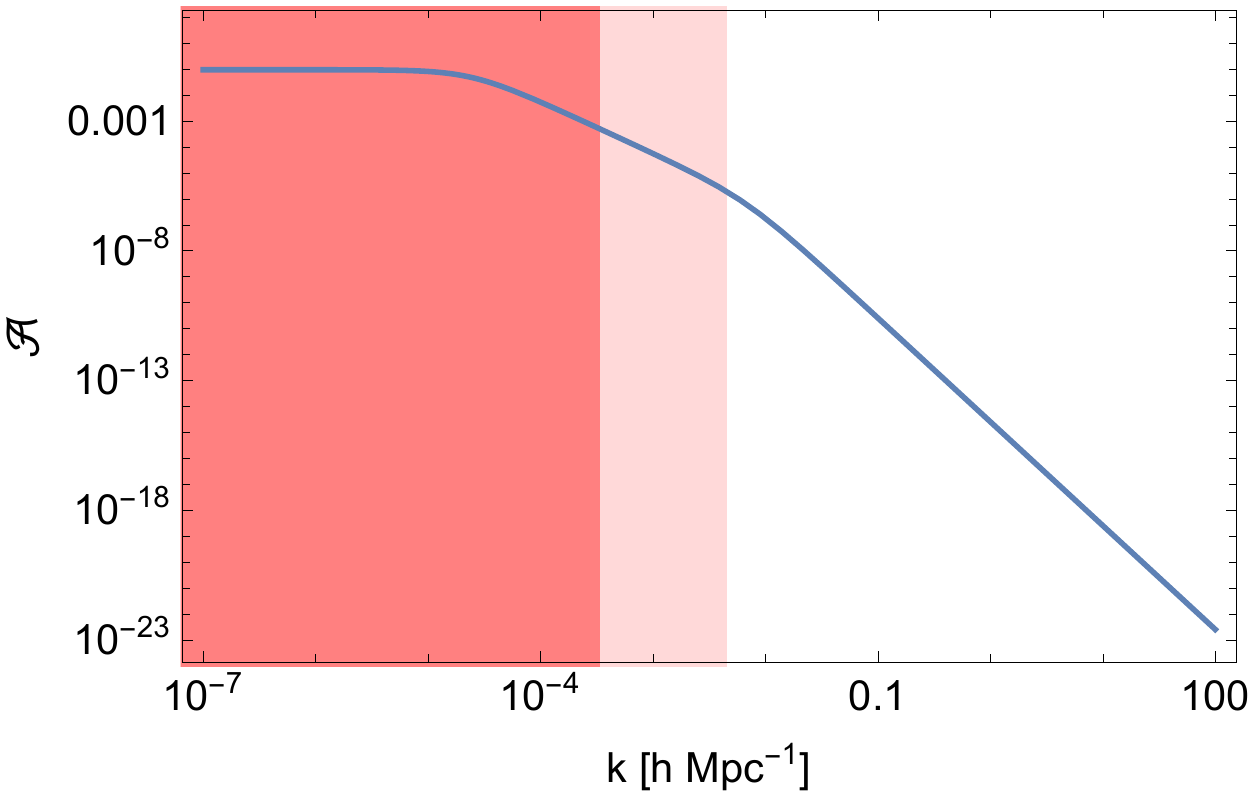}\quad
\includegraphics[width=.3\textwidth]{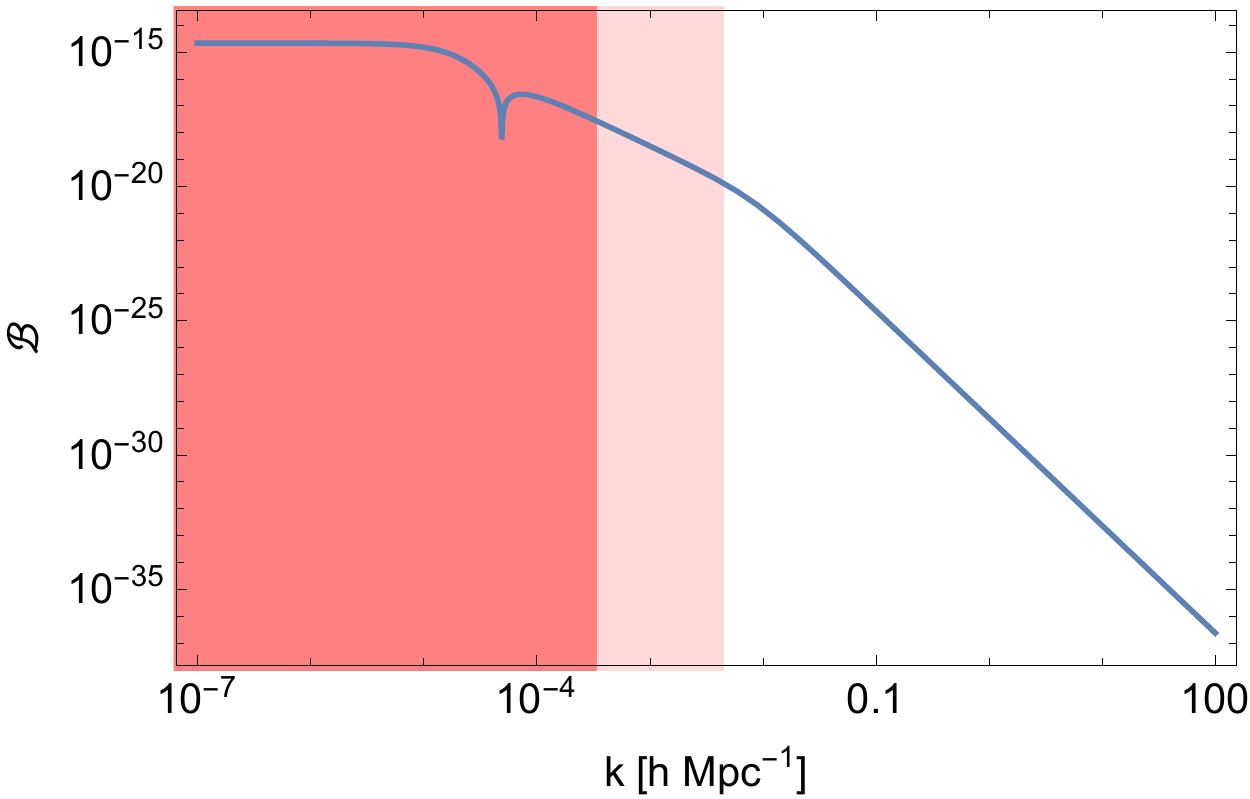}\quad
\includegraphics[width=.3\textwidth]{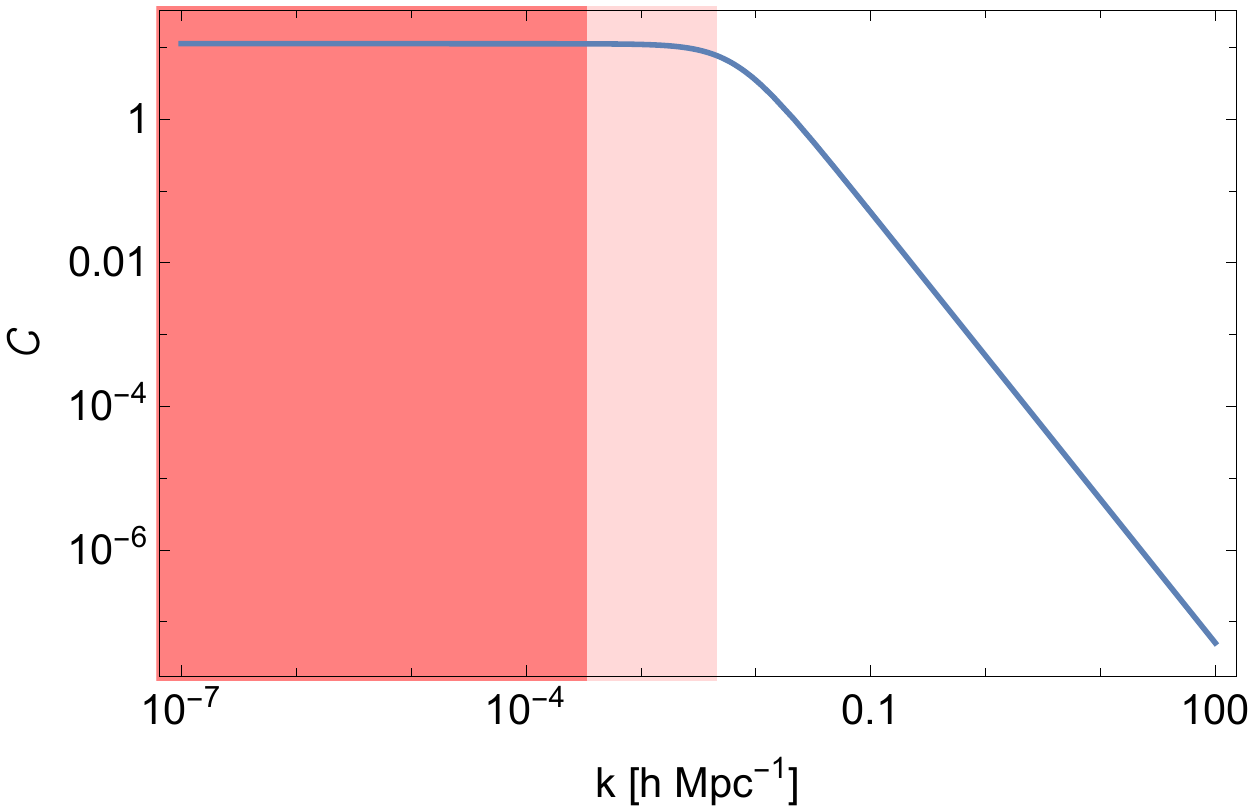}

\medskip

\includegraphics[width=.3\textwidth]{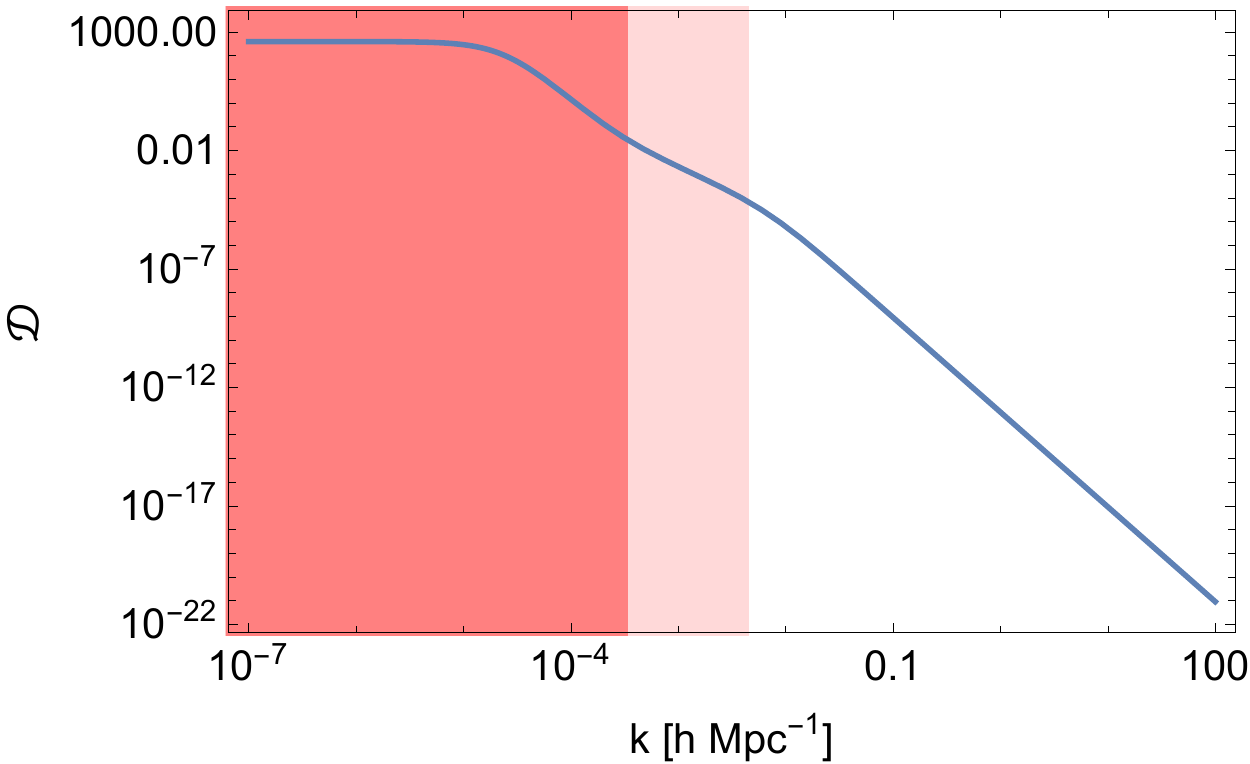}\quad
\includegraphics[width=.3\textwidth]{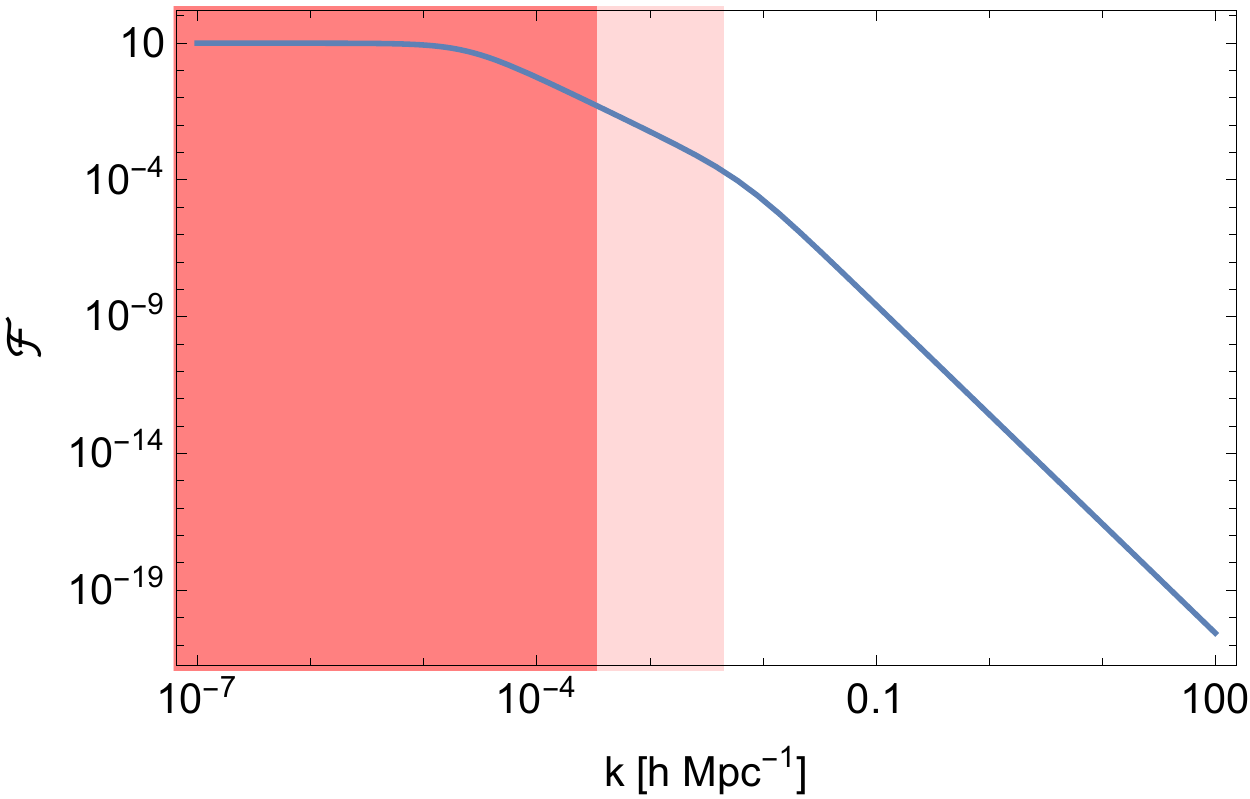}

\caption{Coefficients in (\ref{master}) at $z=0$ as a function of $k$. The left hand side is large scales, whilst the right hand side is small scales. The dark shaded region corresponds super horizon modes with $k < \frac{h}{3000} Mpc^{-1}$. The light shaded region corresponds to modes in the region $\frac{h}{3000} {\rm Mpc}^{-1} < k < k_{*}$, where $k_{*}$ is the wave number corresponding to $\mathcal{E}|_{z=0}=1$. The whole shaded region corresponds to $\mathcal{E}<1$, whilst the area with no shading corresponds to modes with $k> k_{*}$ which represents $\mathcal{E}>1$.
}
\label{coefs}
\end{figure}

With this information, we investigate the coefficients in (\ref{master}) and derive which terms are dominant in the quasi-static approximation (QSA) \cite{Sawicki:2015zya}, as these terms will become relevant when deriving the Poisson's equation. The quasi-static approximation amounts to assuming $\dot{\beta} \sim H \beta \ll \partial_i \beta$
for any perturbation $\beta$,
i.e. we neglect time derivatives with respect to spatial derivatives.
In terms of harmonic modes, it is an expansion for large $k/(aH)$. In Fig.\ref{coefs}, we present a plot of each coefficient at $z=0$ to justify this approximation.

 In Fig.\ref{coefs}, the 3 regimes can be clearly identified. From right to left: 
 regime 1 is $\mathcal{E} \gg 1$, regime 2 (the intermediate regime) is the subhorizon limit with $\mathcal{E} \ll 1$ and regime 3 is the 
superhorizon
 limit. In Appendix \ref{app:masterterms}, we give explicit expressions of the analytic approximations of these functions in each of the 3 regimes. We now apply the QSA to (\ref{master}) as the Poisson's equation is defined in this limit. 
 
 The QSA is valid on scales $10^{-2}\;h\,{\rm Mpc}^{-1}\lesssim k$ while the linear theory is applicable up to $k\lesssim 0.1\; h \,{\rm Mpc}^{-1}$, 
so we consider the regime in which $\mathcal{E} \gg 1$, which corresponds to the non-shaded area of Fig. \ref{coefs}.

\subsection{Poisson's Equation}
In the regime of interest, given by $\mathcal{E} \gg 1$, we calculate the following ratios of the coefficients in ($\ref{master}$) to determine which are the dominant terms in the QSA. We are in essence, comparing the time derivative terms $(\ddot{S},\dot{S})$ to the non-derivative term $S$. Observing that $\tilde{v}\sim \dot{\Delta}/k^2$ from the Euler equation \eqref{eq:allperteqs}, we are also comparing it to the $\Delta$ term:
\begin{equation}\label{case1}
    \left(\frac{H^2}{H_0^2}\right)\frac{\mathcal{A}}{\mathcal{C}}\sim \frac{a^2H^2}{k^2}\mathcal{O}(J) \,,\qquad  \left(\frac{H}{H_0}\right)\frac{\mathcal{B}}{\mathcal{C}} \sim \frac{a^2H^2}{k^2}\mathcal{O}(J) \,,\qquad
     \left(\frac{H_0H}{k^2}\right)\frac{\mathcal{F}}{\mathcal{D}} \sim \frac{a^2H^2}{k^2}.
\end{equation}
From (\ref{case1}), we see that in this regime, ($\ddot{S},\dot{S},\tilde{v})$ can be neglected as their coefficients are suppressed with respect to $(S,\Delta)$ by 
$a^2H^2/k^2$.
In this approximation the master equation (\ref{master}) reduces to the algebraic equation,
\begin{equation}
    H_0^2 \mathcal{C}S+\mathcal{D}\Delta=0,
\end{equation}
for which the solution for $S(\Delta)$ at leading order in $k$ is,
\begin{align}\label{scase1}
    S&=-\frac{\mathcal{D}}{H_0^2 \mathcal{C}}\Delta = \big[2 a^2 J \xi r^2 (J\xi-2\Gamma)\rho\big] \frac{\Delta}{k^2}\; \times \, \nonumber\\ &\Bigg\{2 \kappa M_p^2 J \xi r(2(r-1)\Gamma +J(r+2)\xi)-2 \sqrt{\kappa}M_p^2a J \xi r(H(4(r+1)\Gamma+J \xi(r-4))-2J \dot{\xi})\, \nonumber\\ &+a^2 \Bigg[4M_p^2 H^2\big(2(r+1)\Gamma^2+J \xi(r-4)\Gamma-2J^2 \xi^2(r-1)\big)
    -4M_p^2 H\xi(\Gamma (2(r+1)\dot{J}+J \dot{r})-J(\xi(J\dot{r}-\dot{J}(r-2))+r\dot{\Gamma}))\, \nonumber\\
    &+\xi \big(m^2M_p^2 J^3(2-3r)r \xi^2+2J^2\xi r(m^2M_p^2(3r-1)\Gamma +\rho)+2M_p^2(r+1) \dot{J}^2\xi-2J(r(\Gamma \rho+M_p^2 \xi \ddot{J})-M_p^2 \xi \dot{J}\dot{r})\big)\Bigg]\Bigg\}^{-1}.
\end{align}
Manipulating the equations of motion in Appendix \ref{app:einsteineqs}, we can derive a Poisson's equation for the Newtonian potential sourced by the matter perturbations. In the QSA limit, the equation takes the form:
\begin{equation}\label{poss1}
    \Phi=\frac{1}{4}m^2 a^2 (2\Gamma-J \xi)S+\frac{1}{8r k^2}\Bigg[3m^2 a^4(m^2 J^2 r \xi^2 +H^2(4\Gamma-2J\xi))S+8ra^2\left(m^2\frac{3}{4}\kappa J \xi S-\frac{\Delta\; \rho}{2M_p^2}\right)\Bigg].
\end{equation}
Substituting (\ref{scase1}) into (\ref{poss1}) we obtain the Poisson's equation for the Newtonian potential $\Phi(\Delta)$ in the QSA. 
\begin{equation}\label{possfull}
    \Phi=-\frac{\Delta \rho}{(k^2/a^2)}G(z),
\end{equation}
where $G(z)$ is the scale independent effective Newton's constant, whose value in GR is given by $G_{GR}=\frac{1}{2}M_p^2$. The full Poisson's equation is too complicated to present, so we perform the dRGT expansion after substituting the solution for $S$. Upon substitution, and taking the leading order in dRGT expansion
\footnote{The dRGT expansion is used to obtain a analytic understanding of the system and is not a truly accurate representation of the underlying physics, therefore in the subsequent analysis we give all results without performing the dRGT expansion but still working in the UV limit of the theory to make contact with the QSA.}, we obtain 
\begin{equation}
    \Phi=-\frac{\Delta\; \rho}{2 M_p^2 (k^2/a^2)}\left(1+\frac{m^2 J_1 \xi a}{2 \sqrt{\kappa}H}\right).
\end{equation}
The term in the brackets is the extra factor with respect to GR. Note that in dRGT we have $J_1=0$, so the Poisson's equation is the same as in GR. This is not the case in GMG. 

\subsection{Phenomenology}
In this section we study the phenomenology of the theory. 
Our result will determine whether the fifth force is in effect, ultimately allowing us to determine if there is a need for a screening mechanism on local scales to recover GR.
\begin{figure}[ht]
    \begin{center} 
    \includegraphics[width=0.7\columnwidth]{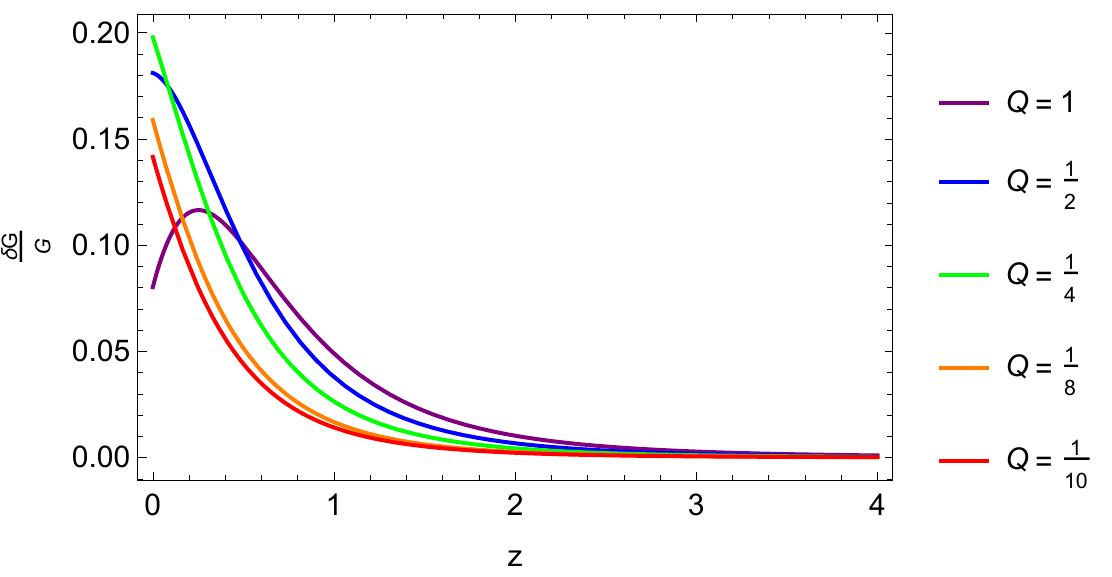}
   \end{center}
    \caption{$\frac{G-G_{GR}}{G_{GR}}$ as a function of redshift for varying values of $Q$.}
    \label{delG}
\end{figure}
Fig. (\ref{delG}) shows a comparison of the effective Newton's constant in our model with respect to GR, for varying values of $Q$. The largest deviation at $z=0$, which is around a $20$\% enhancement, is shown by the $Q=\frac{1}{4}$ curve. The smallest deviation at $z=0$, around an $8$\% enhancement, is the $Q=1$ curve. This result matches the one in Fig. (\ref{varyingq}), the reason why $Q=1$ has the lowest deviation in the effective Newton's constant is because the equation of state has already undergone the turn back to $w=-1$, meaning this branch of the theory is closer to LCDM at $z=0$ than other values of $Q$, so this result is consistent with earlier results found in the background section.

\begin{figure}[ht]
    \begin{center} 
    \includegraphics[width=0.7\columnwidth]{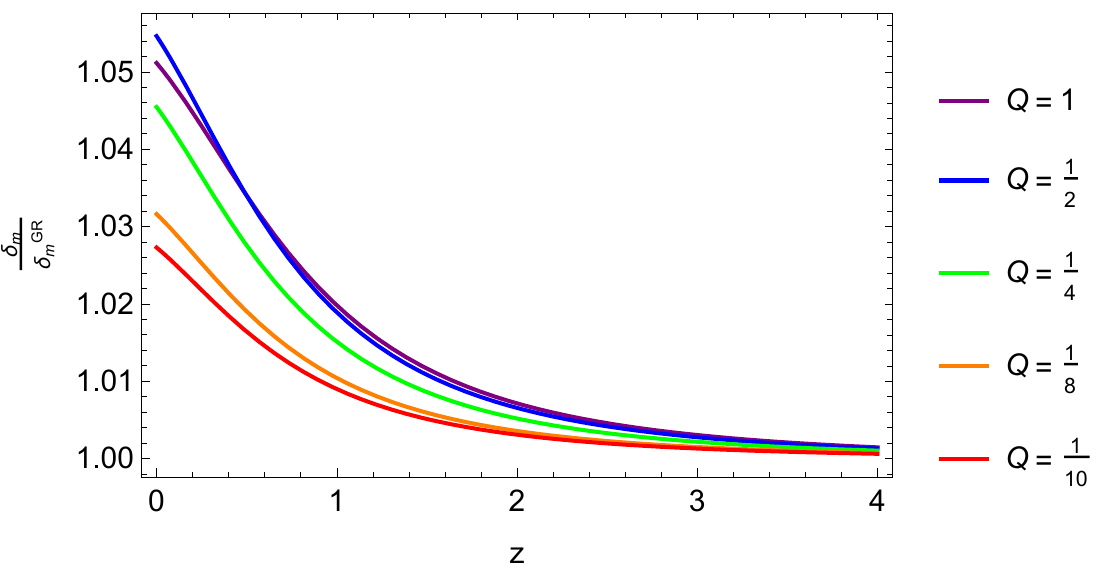}
   \end{center}
    \caption{Ratio of the growth rate in GMG to the growth rate in LCDM as a function of redshift.}
    \label{growthcomp}
\end{figure}

\begin{figure}[h!]
    \begin{center} 
    \includegraphics[width=0.8\columnwidth]{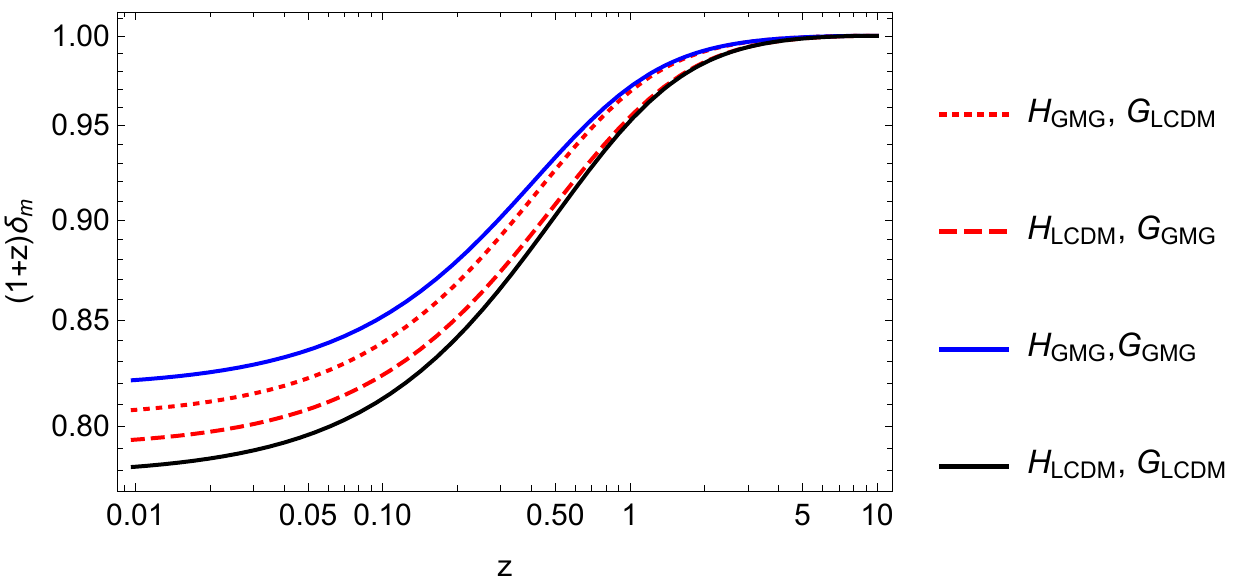}
   \end{center}
    \caption{Growth rate comparison, as a function of redshift, between competing effects in GMG. The black curve shows the result in LCDM. The dashed curve shows result using the Hubble rate in LCDM and the GMG Newton's constant, whilst the dotted curve shows the opposite scenario. The solid blue curve shows the solution using both quantities in GMG.
    }
    \label{comparison}
\end{figure}

A feature of modified gravity models which alter the effective Newton's constant is the modification to the growth rate of matter perturbations. The equation which governs the evolution of the linear matter overdensity, $\delta_m=\Delta/\rho$,
\begin{equation}\label{growth}
    \ddot{\delta}_m+2H\dot{\delta}_m-G\rho \delta_m=0,
\end{equation}
is dependent on two effects. The background expansion $H(z)$ and the effective Newton's constant $G$. As has been seen in earlier section, GMG modifies both of these so one expects the growth rate to be altered with respect to the LCDM case. We take (\ref{growth}) and convert to redshift using the following relations;
\begin{equation}\label{conv}
    \frac{d}{dt}=-(1+z)H\frac{d}{dz}, \qquad \frac{d^2}{dt^2}=(1+z)^2H^2\frac{d^2}{dz^2}+\left[(1+z)^2H\frac{dH}{dz}+(1+z)H^2\right]\frac{d}{dz}.
\end{equation}
Using (\ref{conv}) in (\ref{growth}) we obtain,
\begin{equation}\label{gr1}
    \delta_m''+\left[\frac{H'}{H}-\frac{1}{1+z}\right]\delta_m'-\frac{G \rho}{(1+z)^2H^2}\delta_m=0,
\end{equation}
where a prime denotes a derivative with respect to redshift. We solve (\ref{gr1}) with the initial conditions,
\begin{equation}
    \delta_m(z_i)=\frac{1}{1+z_i}, \qquad \delta_m'(z_i)=-\frac{1}{(1+z_i)^2},
\end{equation}
which are the same initial conditions as LCDM. This is an accurate approximation as GMG mimics the expansion history of LCDM at early times. We solve for $\delta_m(z)$ and compare with the solution in LCDM, the result is shown in Fig. (\ref{growthcomp}).
This time, the highest deviation comes from $Q=\frac{1}{2}$ at around the 5\% level with respect to the LCDM case. To determine the strongest contribution to the growth rate, from the change in the background expansion caused by $H(z)$ or the modification to the Newton's constant, we isolate the $Q=1$ case and solve Eq.~(\ref{gr1}) independently for 3 cases. 
In Fig. (\ref{comparison}), we show a comparison of the full GMG growth function to LCDM, along with two cases where one of $H(z)$ and $G$ are modified.
The main result of Fig. (\ref{comparison}) is that the strongest modification to LCDM occurs when both the effects of the background expansion and the modification to the Newton's constant are included. However, the change in background expansion with respect to LCDM has a larger effect on the growth rate than the modified Newton's constant. This is again consistent with the results of Fig.(\ref{varyingq}), whereby $Q=\frac{1}{2}$ has a larger deviation from LCDM at the background level in $H$ than the case $Q=1$. At early times GMG matches the growth rate of LCDM, however at late times there is a discernible deviation from LCDM.

\subsection{Gravitational wave propagation}
\begin{figure}[h!]
    \centering
    \includegraphics[width=0.8\columnwidth]{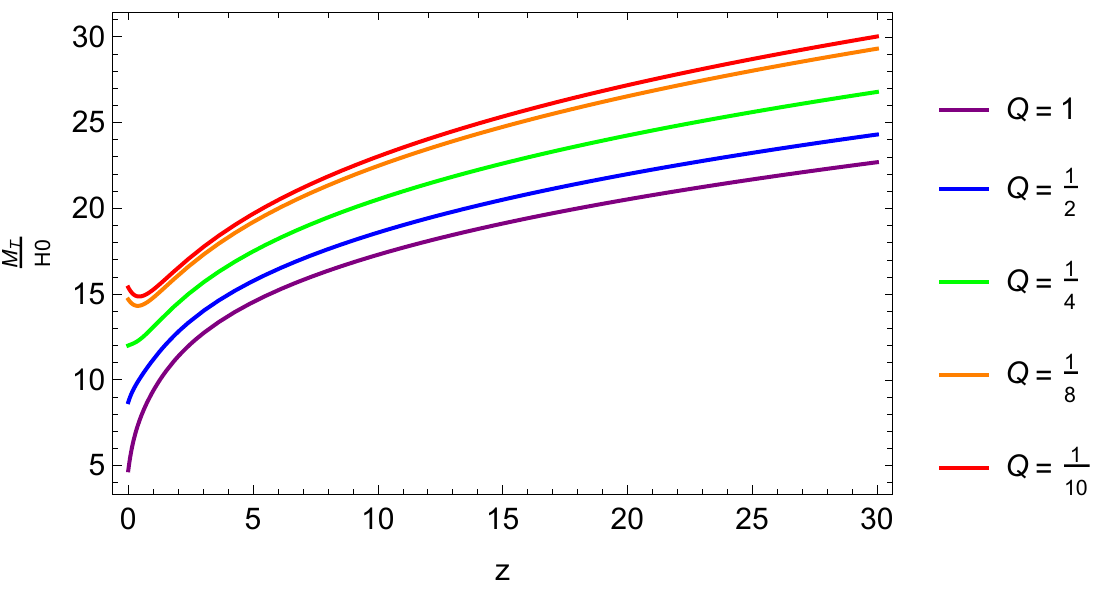}
    \caption{Mass of the tensor modes, normalised to $H_0$ for varying Q.}
    \label{tensormass}
\end{figure}

To study the propagation of gravitational waves, we first quote the equation of motion for the tensor modes \footnote{For the derivation of the quadratic action for the tensors, see Ref. \cite{Kenna-Allison:2019tbu}.}
\begin{equation}\label{tensormodes}
    \ddot{h_{ij}}+3H\dot{h}_{ij}+\left[(1+z)^2(k^2-2\kappa)+M_T^2\right] h_{ij}=0,
\end{equation}
where the only modification with respect to LCDM is that the tensor modes acquire a time dependent mass, $M_T= m \sqrt{\Gamma}$, which is also the same modification as dRGT \cite{Gumrukcuoglu:2012wt}. We plot the tensor mass $M_T$ for varying values of $Q$ shown in FIG. \ref{tensormass}.
For lower values of $Q$, the tensors acquire a higher mass at any given redshift, around an order of magnitude higher than $H_0$. This is due to the normalisation of $\mu$. When we fix $\mu$, imposing that the Hubble rate at $z=0$ matches the value in LCDM, lower values of $Q$ lead to a higher values of $\mu$, therefore the mass of the modes increases towards lower values of $Q$. Additionally, the mass of the tensor modes increases as redshift increases, and this feature is due to $r$ appearing in $\Gamma$ (\ref{defLJ}), which is an increasing function in redshift.

The modification to the mass modifies the speed of gravitational waves. However, the modification is small as $M_T \sim O(10) H_0$ and is well within the bound on the speed of gravity from the detected binary neutron star merger \cite{TheLIGOScientific:2017qsa}. The friction term for the propagation of gravitational waves is unmodified with respect to the LCDM case, which leads to the luminosity distance of gravitational waves in GMG being unmodified with respect to that of light\footnote{Modified gravity models with non-minimal coupling or higher order covariant actions such as the Horndeski theory \cite{Horndeski:1974wa} can produce a modified friction term \cite{Belgacem:2019pkk,Hogg:2020ktc,Belgacem:2018lbp,Belgacem:2017ihm} which can produce a time dependent Newton's constant for gravitational waves and modify the gravitational wave luminosity distance. So it is possible that GMG with non-minimal coupling \cite{Gumrukcuoglu:2020utx} could alter the GW luminosity distance with respect to that of light}.

\section{Discussion and Conclusions}\label{summ}
We considered the Generalised Massive Gravity theory, which is an extension of dRGT theory where the mass parameters are promoted to functions of the four St\"uckelberg fields. For simplicity, we considered a minimal model where all these functions vanish except $\alpha_2$, which slowly varies around its dRGT value and $\alpha_4$, which was chosen to be a non-zero constant compatible with the stability conditions found in Ref.\cite{Kenna-Allison:2019tbu}. Controlling the variation of $\alpha_2$ with the parameter $q = 10^{-4}Q$, we studied the evolution of cosmological solutions in this model. The background is FLRW with an effective fluid corresponding to the mass term, with an equation of state satisfying $w(z)\leq -1$ throughout the evolution. At early times, this effective fluid starts off like a cosmological constant, gradually decreasing. Eventually, it starts to increase again until it reaches $w=-1$ where we lose perturbative control. The time of crossing of the phantom divide can be controlled by the $q$ parameter. We find that for values of $q \lesssim 10^{-4}$, the crossing can be moved to the future. We also find that increasing the amount of negative curvature today has the same effect. The time dependent equation of state modifies the expansion rate by $\mathcal{O}(10^{-2})$ with respect to LCDM, with the maximum deviation at redshifts $z < 1$, depending on the value of $q$.

We also studied the evolution of linear perturbations. We found that the growth function for matter perturbation is modified at $\mathcal{O}(10^{-2})$ with respect to LCDM at low $z$. The non-zero anisotropic stress indicates the presence of a fifth force which contributes to gravitational interactions and increases the effective Newton's constant. This strengthening of gravity contributes to the matter growth, although the modified background evolution $H(z)$ contributes about twice the amount than the former. In the tensor sector, we find that the only modification to gravitational wave propagation arises in the tensor modes picking up a time dependent mass, which increases with redshift and with lower values of $Q$.

The linear study reveals the presence of a fifth force, which needs to be screened at solar system scales. GMG theory is known to admit a Vainshtein mechanism similarly to dRGT \cite{deRham:2014gla}. In order to determine the details of the screening it is necessary to study non-linear perturbations. 
A second motivation for investigating the non-linear behaviour is provided by the dRGT limit of our model. The evolution in the asymptotic past coincides with the self-accelerating branch of dRGT. These solutions have exactly vanishing kinetic terms, controlled by the function $J$, leading to an infinitely strong coupling in the vector and scalar sectors. The solutions in GMG however never have vanishing $J$ although as we go back in the evolution, they do decrease. Whether these modes are strongly coupled depends on the evolution of the non-linear terms and how they depend on $J$. Unlike in dRGT, this is not trivial. The non-linear study will allow us to determine the fate of the perturbative expansion, and provide the scale associated with this strong coupling if it exists.
It should be noted that this issue has been observed in the context of the minimal theory, where only the $\alpha_2(t)$ function varies rapidly. In general, we expect that using a slowly varying function can keep the strong coupling scale finite without affecting the past evolution of the Universe.

Generalised massive gravity is a theory with four arbitrary functions. By using a very limited set of parameters, we have only scratched its surface.
Moreover, GMG has been extended to a general theory class in Ref.~\cite{Gumrukcuoglu:2020utx}, including non-minimal couplings. Finally, relaxing the dRGT constraint to be valid only within the range of the effective field theory, we expect that disformal couplings to matter can be allowed \cite{Gumrukcuoglu:2019rsw}. 
A natural next step is to exploit the full freedom of this theory class, determining new ways to achieve self acceleration and finding other applications in cosmology.

\acknowledgments
MK-A thanks Obinna Umeh for help with xPand \cite{Pitrou:2013hga} which was used to perturb tensorial expressions and Guilherme Brando for useful discussions. The work of AEG and KK has received funding from the European Research Council (ERC) under the European Union’s Horizon 2020 research and innovation programme (grant agreement No. 646702 ”CosTesGrav”). KK is supported by the UK STFC ST/S000550/1. AEG is supported by a Dennis Sciama Fellowship at the University of Portsmouth.

\appendix
\section{Stability Conditions}\label{A}
In this Appendix, we summarise the stability conditions obtained in Ref.\cite{Kenna-Allison:2019tbu}
For the tensor modes, avoiding tachyonic instability requires
    \begin{equation}
  M_T^2= m^2 \Gamma > 0\,.
  \label{eq:stability-tensor}
\end{equation}

For the vector modes, the condition for avoiding ghost and gradient instabilities are, respectively,
\begin{equation}
\frac{m^2 a^2 J \xi}{1+r} > 0 , \qquad 
\frac{(1+r)\Gamma}{2 J \xi} >0\,.
\label{eq:stability-vector}
\end{equation}

Finally, for modes with $k^2 J  \gg a^2 H^2$ (or $\mathcal{E}\gg1$), the scalar graviton is not a ghost if 
\begin{equation} 
\frac{3\,m^2a^4H}{2\,M_p^2r^2}\,\left[
\frac{r\,J\,\xi}{2\,H}\left(\frac{2\,\kappa}{a^2}-\frac{2\,\sqrt{\kappa}\,H}{a} -\frac{4\,H^2}{r}+m^2J\,\xi+\frac{\rho}{M_p^2}\right)+2\,H\,\Gamma-\dot{J}\,\xi
\right] > 0\,.
\label{eq:stability-scalar}
\end{equation}

\section{1st order equations of Motion}
\label{app:einsteineqs}
Using the definitions of gauge invariant variables (\ref{GI1}) then imposing \eqref{BGeqns}, \eqref{stuck} and \eqref{eq:defr}
we obtain the perturbed Einstein equations with one covariant and one contravariant index. For the $0i$ equation, we take out the overall covariant derivative $D_i$. For the traceless part of the $ij$ equation, we remove the overall $D^iD_j - \frac{\delta^i_j}{3}D_lD^l$ operator.  The components are given by
\begin{align}
    \mathcal{E}^{00}&= k^2m^2J\xi S+6m^2aHJr\xi B-\frac{4(k^2+3\kappa)\Psi}{a^2}+a^2 H \Bigg(\frac{6 \tilde{v}\rho }{M_p^2}-3m^2 J r \xi \dot{S}\Bigg)+2\Bigg(\frac{\Delta \rho}{M_p^2}
    +6H^2 \Phi 
    -3m^2 J\xi \Psi-6H \dot{\Psi}\Bigg)\,,\nonumber\\
\mathcal{E}^{0i}&=-2m^2 a J r^2 \xi B-\frac{a^2(r+1)2\tilde{v} \rho}{M_p^2}+m^2 a^2(r+1)J(r-1)\xi \dot{S}-4(r+1)(H\Phi-\dot{\Psi})\,,\nonumber\\
\mathcal{E}^{tr}&= 
-4\frac{(k^2 +6\kappa)\Phi+(k^2+3\kappa)\Psi)}{a^2}
+3m^2 a\Bigg[J r \xi(2\dot{B}-\sqrt{\kappa}(r-1)\dot{S})+2B\Big[H(2\Gamma+J\xi(3r-2))+\xi((r-1)\dot{J}+J \dot{r})\Big]\Bigg]\,\nonumber\\
    &+ 2\Bigg[k^2m^2 \Gamma S+3\sqrt{\kappa}m^2 r (r-1)J\xi B+3m^2(r-2)J\xi\Phi-\frac{6 \rho \Phi}{M_p^2}-6m^2 \Gamma \Psi +6H(3H\Phi+\dot{\Phi}-3\dot{\Psi})\Bigg]  \,\nonumber\\
    &
    - 3m^2 a^2 \Bigg[\big[\xi((r-1)\dot{J}+J\dot{r})+2H(\Gamma+J\xi(2r-1))\big]\dot{S} +r J\xi\ddot{S}\Bigg]
    -12\ddot{\Psi} \,,\nonumber\\
\mathcal{E}^{tl}&=m^2a^2S \Gamma-2(\Phi+\Psi) \,,\nonumber\\
\mathcal{E}^{eu}&= \Bigg[2(k^2+3\kappa)+\frac{3a^2 \rho}{M_p^2}-3a^2(4H^2+m^2 J\xi(r-1))\Bigg]\tilde{v}-2(3a^2 H \dot{\tilde{v}}+\dot{\Delta}+3\dot{\Psi})\,,\nonumber\\
    \mathcal{E}^{co}&= \Phi+a^2(
    2\, H\tilde{v}
    +\dot{\tilde{v}})  \,.
    \label{eq:allperteqs}
\end{align}

\section{Functions in master equation of motion (\ref{master})} 
\label{app:masterterms}
\subsubsection{Case 1: $\mathcal{E} \gg 1$}
\begin{equation*}
    \mathcal{A}=\frac{9\kappa m^2 M_p^2\Gamma_0a^2}{2k^4}, \quad H_0 \mathcal{B}=\frac{9\kappa m^2 M_p^2a^2(\dot{\Gamma_0}+5H\Gamma_0)}{2k^4}, \quad H_0^2\mathcal{C}=\frac{3\sqrt{\kappa}m^2M_p^2\Gamma_0^2aH}{J_1\xi k^2}, \quad \mathcal{D}=\frac{3m^2\Gamma_0a^2 \rho}{2k^4}, \quad  
\end{equation*}
\begin{equation}
    H_0\mathcal{F}=-\frac{27\sqrt{\kappa}m^2 \Gamma_0^2a^5H^2\rho}{J_1\xi k^6}.
\end{equation}
\subsubsection{Case 2: $\mathcal{E} \ll 1$}
\begin{equation*}
    \mathcal{A}=\frac{3J_1\sqrt{\kappa}\xi a m^2 M_p^2 }{4Hk^2}, \qquad H_0\mathcal{B}=\frac{3m^2 \xi \big[2J_1 \kappa^{3/2}M_p^2+2 \kappa M_p^2 a(\dot{J_1}+5J_1 H)+\sqrt{\kappa}a^2 (2\dot{J_1}M_p^2H+8J_1M_p^2H^2+J_1 \rho\big]}{8aH^2k^2}
\end{equation*}
\begin{equation}
    H_0^2\mathcal{C}=\frac{1}{2}m^2 M_p^2 \Gamma_0, \qquad \mathcal{D}=\frac{J_1 \xi m^2 a \rho}{4\sqrt{\kappa}Hk^2}, \qquad H_0\mathcal{F}=\frac{3J_1 \xi m^2 a^3 \rho}{4\sqrt{\kappa}k^2}.
\end{equation}
\subsubsection{Case 3: $k \to 0$}
\begin{equation*}
    \mathcal{A}=\frac{J_1 m^2 M_p^2 a \xi}{4 \sqrt{\kappa}H}, \quad H_0\mathcal{B}=\frac{m^2 \xi \big[2J_1 \kappa M_p^2+2\sqrt{\kappa}M_p^2 a(\dot{J_1}+5J_1H)+a^2(2\dot{J_1}M_p^2 H+8J_1 H^2M_p^2+J_1 \rho)\big] }{8\sqrt{\kappa}aH^2},\quad     H_0^2\mathcal{C}=\frac{1}{2}m^2 M_p^2 \Gamma_0
\end{equation*}
\begin{equation*}
     \mathcal{D}=\frac{m^2\xi \rho}{24 \kappa^{5/2}M_p^2 aH^3}\Big[2J_1 \kappa^2 M_p^2 +2 \kappa^{3/2}M_p^2 a(\dot{J_1}+3HJ_1)+J_1 \sqrt{\kappa}a^3 H\rho+J_1 a^4H^2 \rho +\kappa a^2(2\dot{J_1}M_p^2 H+4J_1M_p^2H^2+J_1\rho)\Big]
\end{equation*}
\begin{equation}
    H_0\mathcal{F}=\frac{J_1\xi m^2 a^3 \rho}{4\kappa^{3/2}}.
\end{equation}

\bibliography{refGM}

\end{document}